\documentclass[sn-nature,Numbered]{sn-jnl}

\usepackage{graphicx}%
\usepackage{multirow}%
\usepackage{amsmath,amssymb,amsfonts}%
\usepackage{amsthm}%
\usepackage{mathrsfs}%
\usepackage[title]{appendix}%
\usepackage{xcolor}%
\usepackage{textcomp}%
\usepackage{manyfoot}%
\usepackage{booktabs}%
\usepackage{algorithm}%
\usepackage{algorithmicx}%
\usepackage{algpseudocode}%
\usepackage{listings}%
\usepackage{comment}%
\usepackage{subcaption}%
\usepackage{booktabs}%
\usepackage{float}
\usepackage{nameref}

\theoremstyle{thmstyleone}%
\theoremstyle{thmstyletwo}%

\theoremstyle{thmstylethree}%

\newcommand{\lya}{Ly$\alpha$}
\newcommand{\fcgs}{erg~s$^{-1}$~cm$^{-2}$~\AA$^{-1}$}

\newcommand{\flcgs}{erg~s$^{-1}$~cm$^{-2}$}
\newcommand{\sblcgs}{erg~s$^{-1}$~cm$^{-2}$~arcsec$^{-2}$}
\newcommand{\msun}{M$_\odot$}
\newcommand\ion[2]{\text{#1\,\textsc{\lowercase{#2}}}} 

\newcommand{\araa}{Annu. Rev. Astron. Astrophys.}   
\newcommand{\aj}{Astron. J.}   
\newcommand{\apj}{Astrophys. J.}   
\newcommand{\apjl}{Astrophys. J. Lett.}   
\newcommand{\apjs}{Astrophys. J. Suppl. Ser.}   
\newcommand{\aap}{Astron. Astrophys.}   
\newcommand{\mnras}{Mon. Not. R. Astron. Soc.}   
\newcommand{\nat}{Nature} 
\newcommand{\pasp}{Publ. Astron. Soc. Pac.}   

\begin{document}

\title[High definition imaging of a $z\approx 3$ filament]{High-definition imaging of a filamentary connection between a close quasar pair at $z =3$}


\author*[1]{\fnm{Davide} \sur{Tornotti}}\email{d.tornotti@campus.unimib.it}
\author*[1,2]{\fnm{Michele} \sur{Fumagalli}}\email{michele.fumagalli@unimib.it}
\author[1,3]{\fnm{Matteo} \sur{Fossati}}\email{matteo.fossati@unimib.it}
\author[1]{\fnm{Alejandro} \sur{Benitez-Llambay}} 
\author[1,4]{\fnm{David} \sur{Izquierdo-Villalba}}
\author[1]{\fnm{Andrea} \sur{Travascio}}
\author[5]{\fnm{Fabrizio}\sur{Arrigoni Battaia}}
\author[1]{\fnm{Sebastiano}\sur{Cantalupo}}
\author[6]{\fnm{Alexander}\sur{Beckett}}
\author[7,8]{\fnm{Silvia}\sur{Bonoli}}
\author[9]{\fnm{Pratika}\sur{Dayal}}
\author[2,10,11]{\fnm{Valentina}\sur{D'Odorico}}
\author[12]{\fnm{Rajeshwari}\sur{Dutta}}
\author[13,14]{\fnm{Elisabeta}\sur{Lusso}}
\author[15,16]{\fnm{Celine}\sur{Peroux}}
\author[6,17]{\fnm{Marc}\sur{Rafelski}}
\author[6]{\fnm{Mitchell}\sur{Revalski}}
\author[18]{\fnm{Daniele}\sur{Spinoso}}
\author[19]{\fnm{Mark}\sur{Swinbank}}

\affil*[1]{\orgdiv{Physics Department}, \orgname{Universit\`a degli Studi di Milano-Bicocca}, \orgaddress{\street{Piazza della Scienza, 3}, \city{Milano}, \postcode{20100}, \country{Italy}}}

\affil[2]{\orgdiv{Osservatorio Astronomico di Trieste}, \orgname{INAF}, \orgaddress{\street{via G. B. Tiepolo 11}, \city{Trieste}, \postcode{34143}, \country{Italy}}}

\affil[3]{\orgdiv{Osservatorio Astronomico di Brera}, \orgname{INAF}, \orgaddress{\street{via Brera 28}, \city{Milano}, \postcode{21021}, \country{Italy}}}

\affil[4]{\orgname{INFN}, \orgdiv{Sezione di Milano-Bicocca} \orgaddress{\street{Piazza della Scienza 3}, \city{20126 Milano}, \country{Italy}}}

\affil[5]{\orgname{Max-Planck-Institut f\"ur Astrophysik}, \orgaddress{\street{Karl-Schwarzschild-Str. 1}, \city{D-85748 Garching bei M\"unchen}, \country{Germany}}}

\affil[6]{\orgdiv{Space Telescope Science Institute}, \orgaddress{\street{3700 San Martin Drive}, \postcode{MD 21218}, \city{Baltimore}, \country{USA}}}

\affil[7]{\orgdiv{Donostia International Physics Center (DIPC)}, \orgaddress{\street{Manuel Lardizabal Ibilbidea 4}, \postcode{E-20018}, \city{San Sebastián}, \country{Spain}}}

\affil[8]{\orgname{IKERBASQUE}, \orgdiv{Basque Foundation for Science}, \orgaddress{\postcode{E-48013}, \city{Bilbao}, \country{Spain}}}

\affil[9]{\orgdiv{Kapteyn Astronomical Institute}, \orgname{Rijksuniversiteit Groningen}, \orgaddress{\street{Landleven 12}, \postcode{9717 AD}, \city{Groningen}, \country{the Netherlands}}}

\affil[10]{\orgdiv{Scuola Normale Superiore}, \orgaddress{\street{P.zza dei Cavalieri}, \postcode{I-56126 Pisa}, \country{Italy}}}

\affil[11]{\orgdiv{Institute for Fundamental Physics of the Universe}, \orgname{IFPU}, \orgaddress{\street{via Beirut 2}, \postcode{I-34151 Trieste}, \country{Italy}}}

\affil[12]{\orgdiv{IUCAA}, \orgaddress{\street{Postbag 4}, \postcode{Pune 411007}, \city{Ganeshkind}, \country{India}}}

\affil[13]{\orgdiv{Dipartimento di Fisica e Astronomia}, \orgname{Università di Firenze}, \orgaddress{\street{via G. Sansone 1}, \postcode{I-50019 Sesto Fiorentino}, \city{Firenze}, \country{Italy}}}

\affil[14]{\orgdiv{Osservatorio Astrofisico di Arcetri}, \orgname{INAF}, \orgaddress{\street{Largo Enrico Fermi 5}, \postcode{I-50125 Firenze}, \country{Italy}}}

\affil[15]{\orgdiv{European Southern Observatory}, \orgaddress{\street{Karl-Schwarzschildstrasse 2}, \postcode{D-85748 Garching bei München}, \country{Germany}}}

\affil[16]{\orgdiv{Aix Marseille Université}, \orgname{CNRS}, \orgaddress{\street{LAM (Laboratoire d’Astrophysique de Marseille) UMR 7326}, \postcode{F-13388 Marseille}, \country{France}}}

\affil[17]{\orgdiv{Department of Physics and Astronomy}, \orgname{Johns Hopkins University}, \orgaddress{\postcode{MD 21218}, \city{Baltimore}, \country{USA}}}

\affil[18]{\orgdiv{Department of Astronomy, MongManWai Building}, \orgname{Tsinghua University}, \orgaddress{\city{Beijing 100084}, \country{People’s Republic of China}}}

\affil[19]{\orgdiv{Centre for Extragalactic Astronomy}, \orgname{Department of Physics, Durham University}, \orgaddress{\street{South Road}, \city{Durham DH1 3LE}, \country{UK}}}


\abstract{\textbf{Filaments connecting halos are a long-standing prediction of cold dark matter theories. We present a detection of the cosmic web emission connecting two quasar-host galaxies at redshift $z \sim 3.22$ in the MUSE Ultra Deep Field (MUDF), observed with the Multi Unit Spectroscopic Explorer (MUSE) instrument. The very deep observations unlock a high-definition view of the filament morphology, a measure of the transition radius between the intergalactic and circumgalactic medium, and the characterization of the surface brightness profiles along the filament and in the transverse direction. Through systematic comparisons with simulations, we validate the filaments' typical density predicted in the current cold dark-matter model. Our analysis of the MUDF field, an excellent laboratory for quantitatively studying filaments in emission, opens a new avenue to constrain the physical properties of the cosmic web and to trace the distribution of dark matter on large scales.}}

\maketitle


\section*{Main Text}\label{main}
The existence of cosmic filaments connecting halos hosting galaxies has been a long-standing prediction of theories describing a dark-matter dominated Universe. Already from earlier comparisons between N-body simulations and galaxy surveys, it became clear that models including pancake-like structures were superior in reproducing the observed galaxy distribution, hinting at the fact that galaxies trace an underlying mass distribution that extends beyond a few Mpc \cite{Frenk1988}. Further development of simulations including baryons \cite{Cen1994,Lukic2015}, the clustering analysis in ever-growing galaxy redshift surveys \cite{Peacock2001,Tempel2014}, and the ability of quasar spectroscopy to map the shadows of diffuse gas in absorption \cite{Rauch1998} have contributed to shaping our view of the intergalactic medium (IGM) as composed of a cosmic web: a network of filaments extending on Mpc-scales at the intersection of which dark matter overdensities become the cradles where gas collapses and forms galaxies. 

Direct imaging of these filaments has proven challenging for several decades, as theoretical and numerical works predict that the filaments emit fluorescence radiation induced by ultraviolet background (UVB). Infact, the low intensity of the UVB \cite{Haardt&Madau2012} leads to an expected surface brightness emission of $\approx 10^{-20}$ \sblcgs~ at redshift $z\sim 3$ \cite{Gould&Weinberg1996}, below the sensitivity limits of previous instruments. The deployment of large-format integral field spectrographs, such as the Multi Unit Spectroscopic Explorer (MUSE) \cite{Bacon2010MUSE} at the Very Large Telescope (VLT) and the Keck Cosmic Web Imager (KCWI) \cite{Morrissey2018}, has marked a breakthrough in studying the low-surface-brightness Universe. Mapping gas around local ionizing sources, such as quasars, has become a routine experiment with multiple examples of $\gg 10^{-18}$ \sblcgs\ \lya\ nebulae, typically on scales of a few hundred physical kiloparsecs (pkpc), known \cite{Martin2014a, Husband2015,Borisova2016,ArrigoniBattaia2019,Fossati2021}.
Among these examples, features approaching the megaparsec scale, as in the Slug Nebula \cite{Cantalupo2014}, provided first hints of filaments. More recently, the enhanced sensitivity has allowed us to obtain the first images of patches of ionized gas stretching over scales of the order of $\approx 1~$physical Mpc in a $z\approx 3.1$ galaxy protocluster \cite{Umehata2019}, and to identify filamentary emission connecting galaxies \cite{Bacon2021}. \lya\ emission from structures similar to bridges has also been observed around active galactic nuclei (AGN) \cite{Cai2018, ArrigoniBattaia2019b, Herenz2020}, and statistically detected in the intergalactic medium \cite{Martin2023}.

This work presents a detection and quantitative characterization of a cosmic web filament through its emission, at a surface brightness of $\approx 8 \times 10^{-20}$ \sblcgs, connecting two massive halos hosting quasars at $z\approx 3.22$ in the MUSE Ultra Deep Field (MUDF, \cite{Revalski2023, Fossati2019}). The brighter quasar J2142-4420 has a continuum AB magnitude, $m_\mathrm{r} = 17.9 \pm 0.02$ and systemic redshift $z=3.221\pm0.004$; the fainter quasar J2142-4419 has $m_\mathrm{r} = 20.5 \pm 0.03$ and systemic redshift $z=3.229\pm0.003$ (\cite{Lusso2019}).
The MUSE data, totaling 142 hours on-source, allow us to image in high definition the entire emitting structure, which stretches for $\sim 700$ pkpc between and sideways of the two halos. 
The data quality enables a detailed investigation of the \lya\ emission from the IGM for more than $250$~pkpc beyond the virial radii, at the low surface brightness predicted for cosmic filaments. With this data we were able to map the \lya\ surface brightness profile along the filament's spine and in the transverse direction. Finally, comparisons with numerical simulations offer insight into the typical density of cosmic filaments, a main prediction of the current cold dark-matter models.

\begin{figure}
\centering
\includegraphics[scale=0.5]{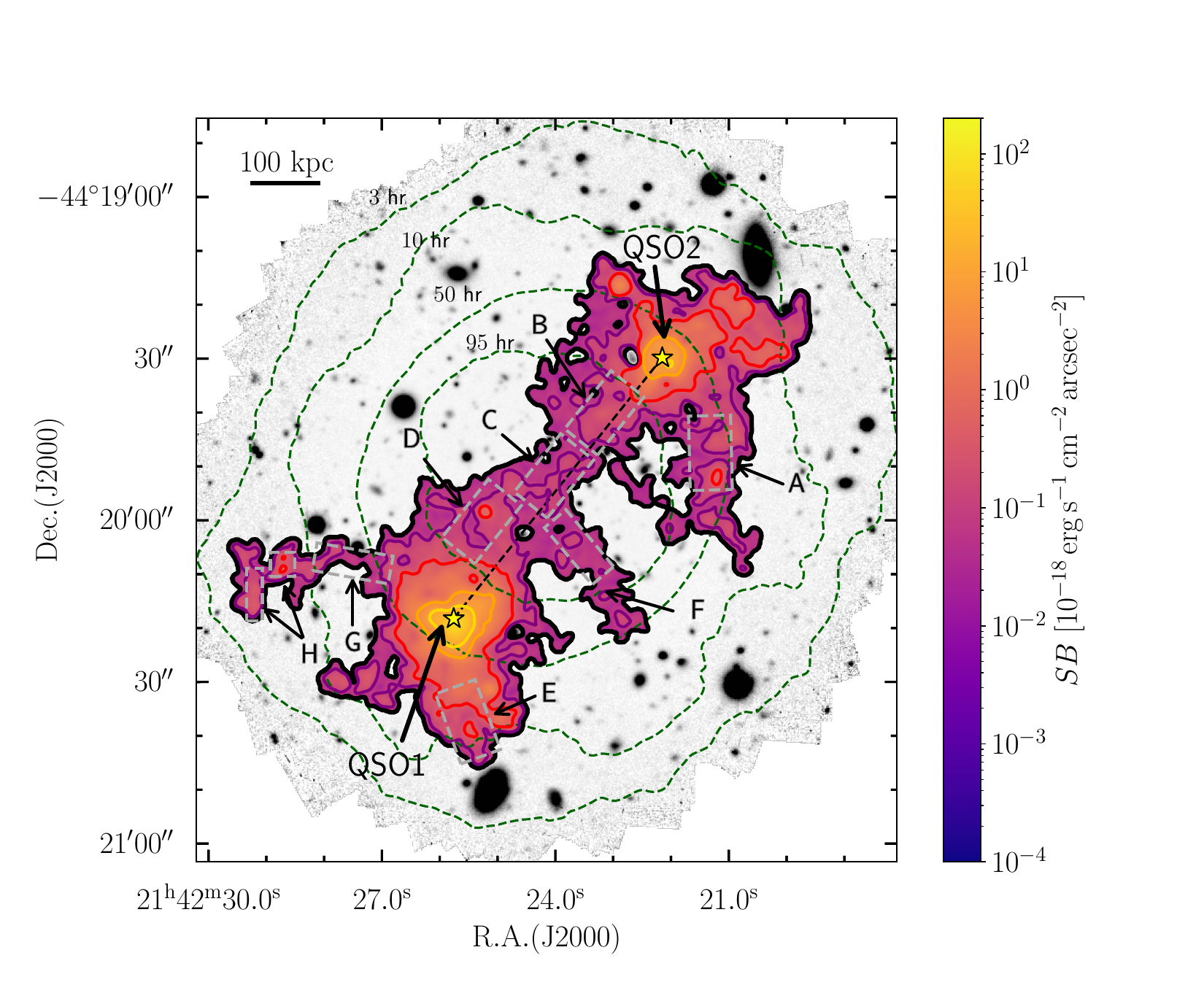}
\caption{\textbf{\lya~image of the filament in the MUDF.} Optimally-extracted \lya\ image of the extended nebulae surrounding the two quasars (marked by yellow stars and the respective labels QSO1 and QSO2) and of the filament connecting them along the diagonal direction (dashed black line, $\approx 500$~pkpc). The contour levels are $0.02$, $0.1$, $0.4$, $3.2$, and $10 \times 10^{-18}$~\sblcgs~ (black, purple, red, orange, and gold). The black contour is the detection limit at $S/N = 2$. The color bar covers the same values of Fig. \ref{fig:4}. The dashed grey boxes labeled from A to H are the same in Extended Data Figure \ref{fig:ext1}. The background, in grey, is a white-light image of the region imaged by MUSE, with the dashed green contours defining the exposure time map of the field according to the labeled values.}
\label{fig:1}
\end{figure}

Prior analysis of a partial dataset of $\approx 40$~hr \cite{Lusso2019} uncovered extended \lya\ nebulae in the circumgalactic medium (CGM) of the quasar hosts, with asymmetric extensions along the direction of the two active galactic nuclei. Based on these features that indicated a gaseous bridge, we searched the region across the two quasars for low-surface brightness \lya\ emission by selecting groups of connected pixels with a signal-to-noise ratio ($S/N$) above a threshold of 2 (see Methods, for further details). This exercise identified \lya\ emission in a connected region of $\approx 30\times 90$~arcsec$^2$.
Fig. \ref{fig:1} shows the optimally projected surface brightness map of the connected emission, along with labels for the two quasars, QSO1 (the brighter) and QSO2 (the fainter). Table~\ref{tab:global_properties} summarizes the main properties of the emitting structures. 

\begin{table}
    \centering
    \begin{tabular}{ccccc}
        \toprule
           & Area & Mean Surface Brightness  & Integrated flux&  Size \\
               & (arcsec$^2$) & ($10^{-19}$~erg s$^{-1}$ cm$^{-2}$ arcsec$^{-2}$) & ($10^{-16}$ erg s$^{-1}$ cm$^{-2}$) & (pkpc) \\
        \midrule
        Nebula 1 & $635^*$ & $17.4 \pm 0.1$ & $11.1 \pm 0.1^*$  & 117 \\
        Nebula 2 & $530^*$ & $7.1 \pm 0.1$ & $3.78 \pm 0.05^*$ & 108 \\
        Filament & $830^*$ & $0.83\pm 0.06$ & $0.69 \pm 0.05^*$ & 250 \\
        \bottomrule
    \end{tabular}
    \caption{\textbf{Global \lya\ properties of the two nebulae and the filament.} The nebulae properties are computed using circular apertures up to the transition radius between the CGM and IGM and centered on the quasar. This transition radius also defines the size of the nebulae. The properties of the filament are calculated using a box between the two nebulae, at a distance given by the transition radii. The length of this box defines the reported size of the filament. $^*$These quantities depend on the selected analysis region (see Extended Data Figure \ref{fig:ext2}).}
    \label{tab:global_properties}
\end{table}

An extended emission stretches for over $\approx 700$~pkpc in projection, both in the opposite and the in-between directions of the two quasars. With a mean surface brightness of $\approx 8 \times 10^{-20}$~\sblcgs, the filamentary structure between the two quasars would not have been detected in shallower data acquired by most surveys reaching $\gg 10^{-19}$~\sblcgs. 
We confirm the emission by extracting spectra from different regions along the main filament (boxes B, C, and D in Fig. \ref{fig:1} and Extended Data Figure \ref{fig:ext1}). The map also reveals protuberances and smaller sub-structures branching from the main filament. 
Some of these substructures lie in regions of moderate depth due to the non-uniform sensitivity of the map (see, e.g., box G in Fig. \ref{fig:1} and Extended Data Figure \ref{fig:ext1}), but the two largest protuberances — one extending sideways from the central part of the main filament to the west (box F), and the other from the nebula of QSO2 to the southwest (box A) — are spectroscopically confirmed.
The general morphology of the system, composed of galaxies, nebulae, and filaments, is remarkably similar to the configuration of galaxies assembling inside the cosmic web predicted by modern cosmological simulations \cite{Rosdahl&Blaizot2012,Rahmati2015}. Direct evidence of narrow filaments protruding from halos is also reminiscent of the cold-mode accretion proposed by numerical simulations (\cite{Keres2005, Dekel2009}), as also seen in previous observations (\cite{Rauch2016, Daddi2021}).

\begin{figure}
  \begin{minipage}{1\textwidth}
    \centering
    \begin{subfigure}{0.8\textwidth}
        \includegraphics[width=\textwidth]{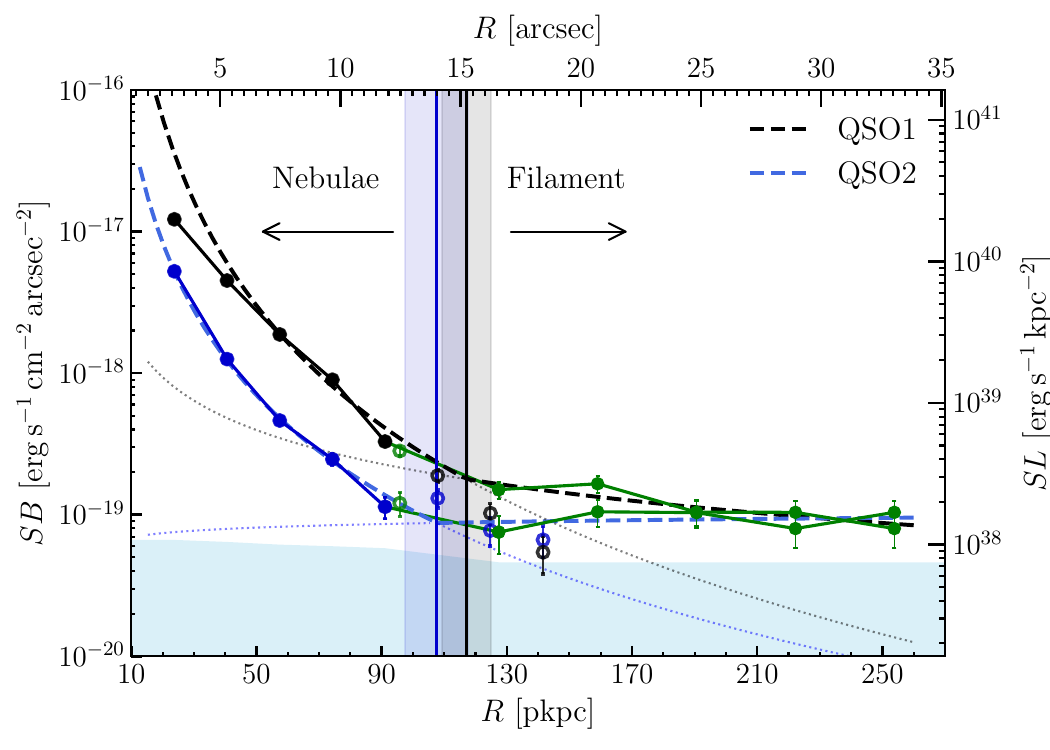}
        \caption{}
        \label{fig:2a}
    \end{subfigure}
  \end{minipage}\hfill
  \begin{minipage}{1\textwidth}
    \centering
    \begin{subfigure}{0.8\textwidth}
        \includegraphics[width=\textwidth]{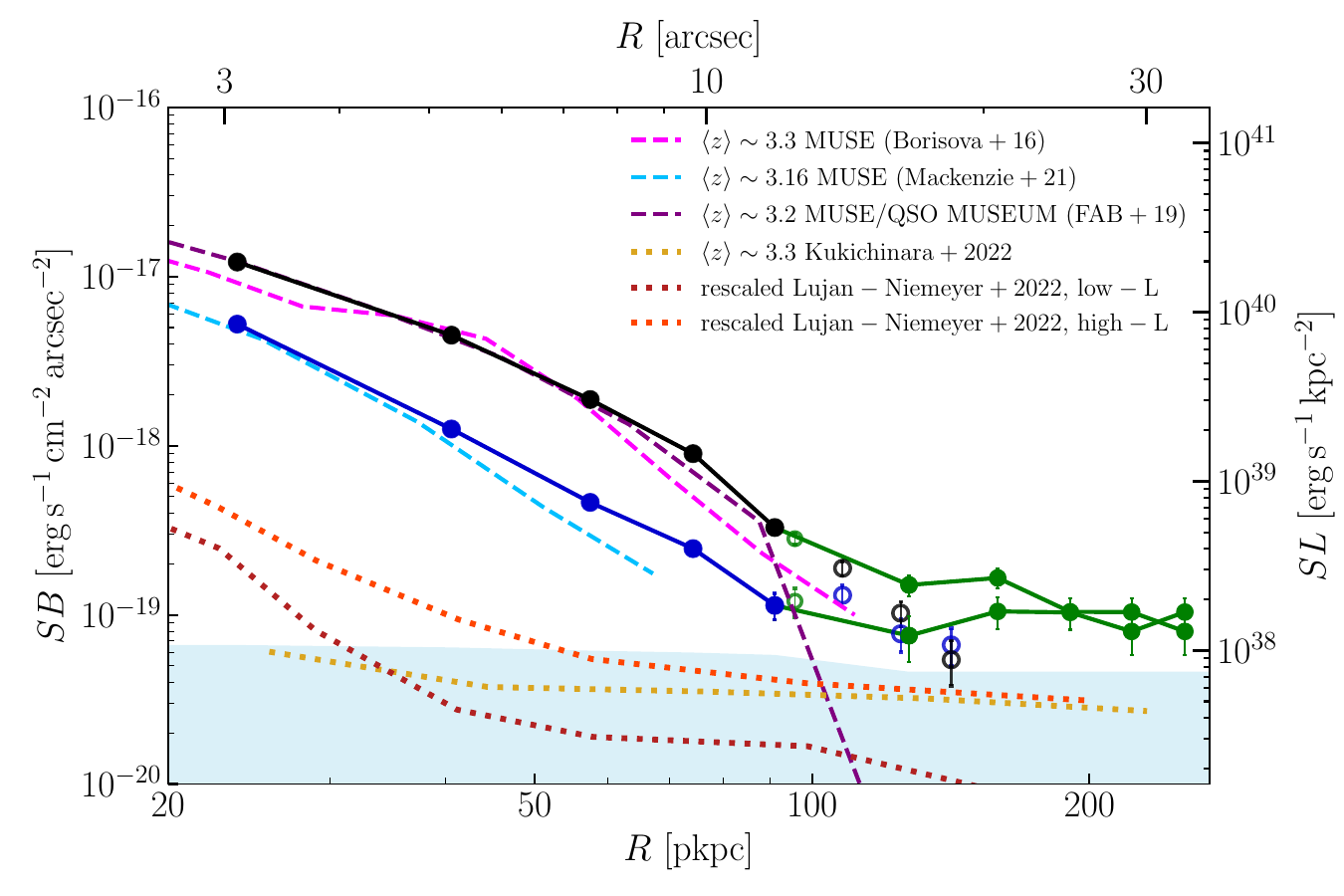}
        \caption{}
        \label{fig:2b}
    \end{subfigure}
  \end{minipage}
  \caption{\textbf{The \lya\ surface brightness profiles of the nebulae and filament.} \textbf{(a)} The extended \lya\ emission's surface brightness at the positions of QSO1 (black solid line) and QSO2 (blue solid line) are shown with filled dots, along with the filament profile (green solid line). The empty dots are the remaining data points measured using the different apertures (see Methods).  All points represent weighted averages along with their standard errors. The dashed lines (black and blue) represent the best-fit double power-law models. The two vertical solid lines (black and blue) denote the median values, along with the $16^{\rm th}$ and $84^{\rm th}$ percentiles, of the transition radii that separate the quasars' CGM from the filament. The dotted black and blue lines represent the extrapolation of the single power laws of the fitted function. \textbf{(b)} Comparison between the measured surface brightness profiles shown as in (a) but in log-log scale and literature profiles (dashed lines) in samples at comparable redshift and quasar magnitudes (see the text).  
The dotted lines represent stacked profiles around LAEs. The horizontal sky-blue region in both figures marks the narrow-band detection limit (see Methods for the definition).}
  \label{fig:2}
\end{figure}

The depth of the MUSE observations in the MUDF provides sufficient data to analyze the structural properties of the filament. Starting with the flux-conserving projected surface brightness map obtained by collapsing the MUSE datacube in a wavelength window of $30$~\AA\ centered on the wavelength at the redshift of the \lya\ emission peak, we extract the \lya\ surface brightness profile along the axis connecting the two quasars. We adopt a circular geometry and derive the azimuthally average profile for the emission arising in the nebula near the quasars. For the filament emission, we extract instead the average surface brightness along the axis connecting the quasars inside rectangular regions (see Extended Data Figure \ref{fig:ext2} in the Methods for details). The central region within $\approx 15$~kpc of each quasar is excluded from the analysis due to the residual of the quasar point spread functions, which have been subtracted from the original cube. The emission extracted from the nebulae and the filament profile join smoothly, and we use the radius at which they intersect to switch from one geometry to another (see Methods for a detailed explanation). 

The resulting profiles are shown in Fig. \ref{fig:2a} on a linear scale and in Fig. \ref{fig:2b} on a logarithmic scale to highlight better the profile of the quasar nebulae at small radii. The profiles rapidly decline with radius, following a power law with index $\approx -3.3$ for Nebula 1 and $\approx -2.7$ for Nebula 2 (see Methods for details on the fitting procedure) before reaching a plateau of almost constant surface brightness in the region dominated by the filament emission. The profile near QSO2 appears to be a scaled-down version of that centered on QSO1. As shown in the Fig. \ref{fig:2b}, the radial profile measured for the two nebulae is entirely consistent in shape and normalization with the average profiles obtained for bright (median \textit{i}-band magnitude 17.81 in \cite{Borisova2016} and 18.02 in \cite{ArrigoniBattaia2019}) and faint quasar samples (median \textit{i}-band magnitude 21.32 in \cite{Mackenzie2021}). Hence, the MUDF quasar nebulae are typical under this metric.

A slope change is apparent at larger radii when the profiles reach a surface brightness of $\lesssim 10^{-19}$~\sblcgs. We attribute this variation in the power-law index to the transition between the regime dominated by the CGM of the quasar host galaxies and the IGM. This is similar to the analysis performed in a nearby galaxy by Nielsen et al. \cite{Nielsen2023} to infer the transition between the interstellar and circumgalactic medium. By modeling the full profile with a double power law (see Equation~\ref{eq:pwl} in the Methods), we constrain the transition radius between these two regions, $R_t$. For QSO1, $R_t = 117\pm8$~pkpc, while for the fainter quasar QSO2 $R_t = 107 \pm 10$~pkpc (see Table \ref{tab:best_fit_params} in Methods). Values of $\approx 100$~pkpc are comparable to the virial radius of halos with mass $\approx 2-3 \times 10^{12}$~M$_\odot$, which is the estimated halo mass of $z\approx 3$ quasars at these luminosites \cite{Fossati2021,deBeer2023}. Our analysis, therefore, provides one of the very few examples currently available in which the transition radius between the CGM and IGM is directly measured, and the only such measurement in emission at $z\sim3$.
For comparison, at lower redshift ($z<0.5$), using absorption line statistics and by measuring the covering fraction of \ion{H}{I} absorption, Wilde et al. \cite{Wilde2023} found a typical size of about twice the virial radius for the CGM of star-forming galaxies. 

We further compare the surface brightness of the MUDF filament with the radial profiles in LAE stacks by \cite{Kikuchihara2022} and \cite{LujanNiemeyer2022} (dotted lines). Flattening at large radii is also evident in these cases (Fig. \ref{fig:2b}), which can also be interpreted as the halo transition radius, which appears at smaller radii ($40-60$ kpc). However, in stacks, it is more challenging to disentangle the contribution of diffuse gas emission from filaments and the overlapping signal of additional halos. Moreover, we observe a difference in surface brightness by a factor of up to ten and even greater if we consider the SB levels $< 10^{-20}$~\sblcgs\ reached in the ultra-deep stacked profile of \cite{Guo2024} at radii $\gtrsim 50$~kpc. These differences could be explained by the different halo mass scales investigated ($10^{12.5} M_\odot$ for quasars and $10^{10-11} M_\odot $ for LAEs) and by a different ionizing field. Moreover, the stacking technique introduces signal dilution due to geometric effects. Thus, our study provides a complementary and more direct view of filaments at the mass scale of $10^{12.5} M_\odot$.

\begin{figure}
\centering
\includegraphics[scale=0.60]{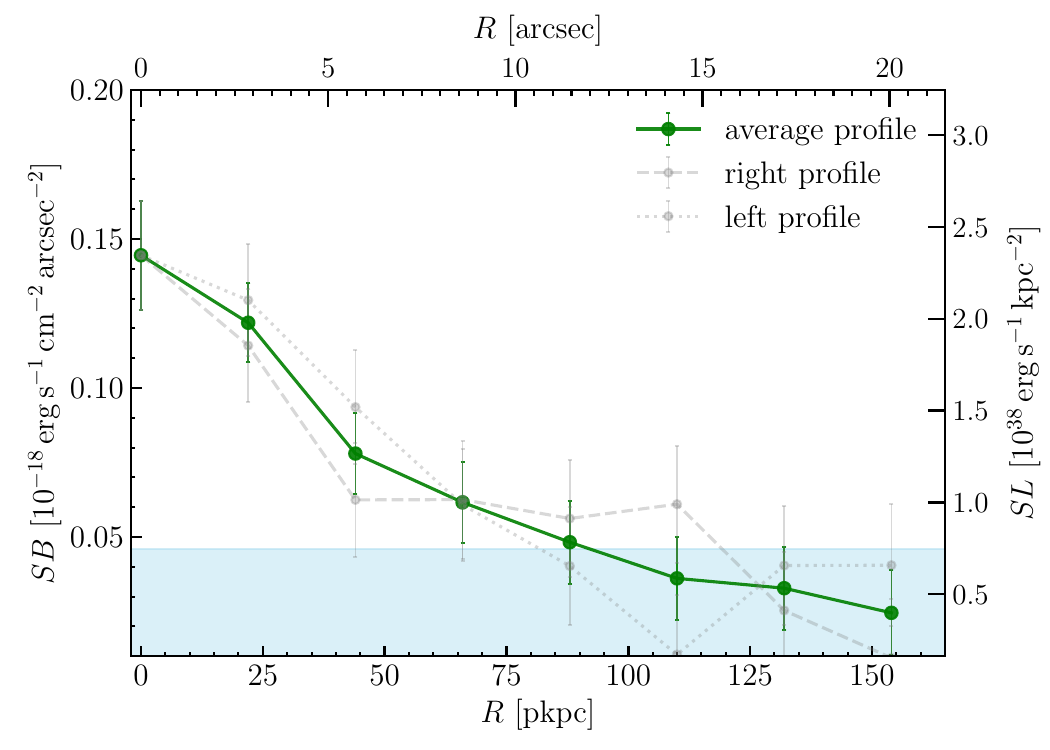}
\caption{\textbf{The transverse \lya\ surface brightness profile of the filament.} The dashed and dotted grey lines represent the right and left profiles relative to the direction connecting QSO1 with QSO2. The green solid line shows the resulting average profile combining the two. All points represent weighted averages along with their standard errors. The horizontal sky-blue region marks the narrow band detection limit (see Methods for the definition).}
\label{fig:3}
\end{figure}

The quality of the MUDF data further allows for the measurement of the filament properties in the direction transverse to that connecting the quasars (i.e., the spine). For this analysis, we employ rectangular extraction regions outside the quasar CGM ($R>R_t$) and measure the average surface brightness profile on both sides of the filament's spine (see Extended Data Figure \ref{fig:ext2} in the Methods). Both sides produce a comparable measure that we average to construct a final transverse \lya\ profile (Fig. \ref{fig:3}). In the transverse direction, the profile drops with a power-law index of $\approx -0.74$ up to $\approx 70$~pkpc. The total thickness is $\lesssim 2\times 70 \approx 140$~pkpc at the depth of our observations. There is no strong evidence in the current data of a clear edge or a change in the profile slope.

Finally, the observed surface brightness is indicative of low-density gas ($n_H \lesssim 10^{-2}~\rm cm^{-3}$) inside the filaments. Adopting simple scaling relations (see, e.g., \cite{ArrigoniBattaia2016}) for a denser optically thick medium, we would obtain surface brightness levels two orders of magnitude higher than those observed, given the quasar luminosities and the distance at which the filament lies. In contrast, the quadratic dependence of the emissivity in an optically thin recombination scenario (see Methods), puts the characteristic density of the filament at $\approx 5\times 10^{-3}$ cm$^{-3}$ for the mean observed surface brightness. However, in light of additional radiative processes and a density distribution within the emitting medium, a more refined inference on the gas density requires modeling of hydrodynamic simulations, as done in the next paragraphs.

The MUDF has been selected for observations because of a pair of bright close quasars. To understand if the properties derived from the MUDF filament can be generalized to other cosmic environments, we assess how typical this double-quasar system is expected to be in a cold dark matter Universe.

For this task, we search for MUDF twins in the semianalytic model (SAM) {\tt L-Galaxies} based on the {\tt Millennium} simulation \cite{Izquierdo-Villalba2019,Izquierdo-Villalba2020} (see Methods for further details). We opt for this model as it implements detailed quasar physics that successfully reproduces key statistics of the quasar population, including the observed luminosity functions reported in the literature. 
By selecting simulated quasar pairs within 0.3~dex of the observed bolometric luminosities ($\approx 2\times 10^{47}$ erg s$^{-1}$ for QSO1 and $\approx 2\times 10^{46}$ erg s$^{-1}$ for QSO2, \cite{Lusso2023}) with projected distances of $400-600$~pkpc and line-of-sight velocity separation of $\Delta v \leq 1000$~km~s$^{-1}$ (as estimated from the redshift of the quasars, see Methods), we find that the three-dimensional physical distance of pairs is closer than 2.5~pMpc for 95\% of the systems, and less than 1~pMpc in half of the cases (see Extended Data Figure \ref{fig:ext3}). 

With this SAM, we also infer a number density of quasar pairs of $5.6\times 10^{-9}$~cMpc$^{-3}$, implying an expected occurrence of one MUDF-like quasar pair within a volume of ($560$~cMpc)$^3$. Using these twins, we derive the distribution of halo masses of bright and faint quasars (see Extended Data Figure \ref{fig:ext4} in the Methods), from which we infer a typical halo mass of $\log(M_\mathrm{vir}/M_\odot) = 12.91^{+0.34}_{-0.33}$ for the brighter quasar and $\log(M_\mathrm{vir}/M_\odot) = 12.25^{+0.46}_{-0.35}$ for the fainter one. These halo mass values are consistent with current estimates of quasar hosts at comparable redshifts \cite{Fossati2021,deBeer2023}.
As we have no reason to expect that the underlying total hydrogen density distribution within the filament depends critically on the presence of quasars, we can use the halo mass distributions obtained from the SAM to select pairs with separations of $\lesssim 1$~pMpc and investigate further with cosmological hydrodynamic simulations how filaments connect them.

For this analysis, we use the intermediate resolution simulation publicly available from the IllustrisTNG project (TNG100$-$1, see Methods; \cite{Nelson2019}), representing a compromise between resolution and volume. Within this box, we identified 144 pairs with halo masses given by the SAM, with a projected distance in the range of $400-600$~pkpc and a 3D distance below $1$~pMpc. Firstly, we observe that these pairs are generally physically connected by a dense filamentary structure in contrast to pairs that have larger physical distances, $>2$~pMpc. 
A thorough inspection of the simulated pairs reveals that the median density profile along the direction that connects the two halos (see Methods) declines from the central regions smoothly reaching a plateau with a minimum value of $\approx 10^{-3.7}$~cm$^{-3}$, i.e., more than ten times the average cosmic density at these redshifts ($\approx 1.4\times 10^{-5}$~cm$^{-3}$). In contrast, for pairs with a physical distance $>2$ pMpc, the minimum value of the median profile reaches $\sim 10^{-4.8}$~cm$^{-3}$ (see Extended Data Figure \ref{fig:ext5}), not far from the mean density. Along the filament's axis for connected pairs, the hydrogen density reaches a median peak value of $10^{-2.8}$~cm$^{-3}$ (see the transverse profile in Extended Data Figure \ref{fig:ext6}), indicating the presence of a denser spine in the central part of the cosmic web. Given these properties, we infer that a typical filament of uniform hydrogen density $\approx 10^{-3}$ cm$^{-3}$ and a radius of $\approx 75$ pkpc would have a total hydrogen column density of $\approx 3-5 \times 10^{20}$ cm$^{-2}$. 
Such a low-density gas exposed to ionizing radiation is predicted to have a neutral fraction $\lesssim 10^{-4}$. We expect a neutral hydrogen column density below those of Lyman-limit systems ($N_\ion{H}{I}<10^{17.2} \rm~cm^{-2}$), which is expected for absorption line systems in the IGM.

\begin{figure}
\centering
\includegraphics[scale=0.35]{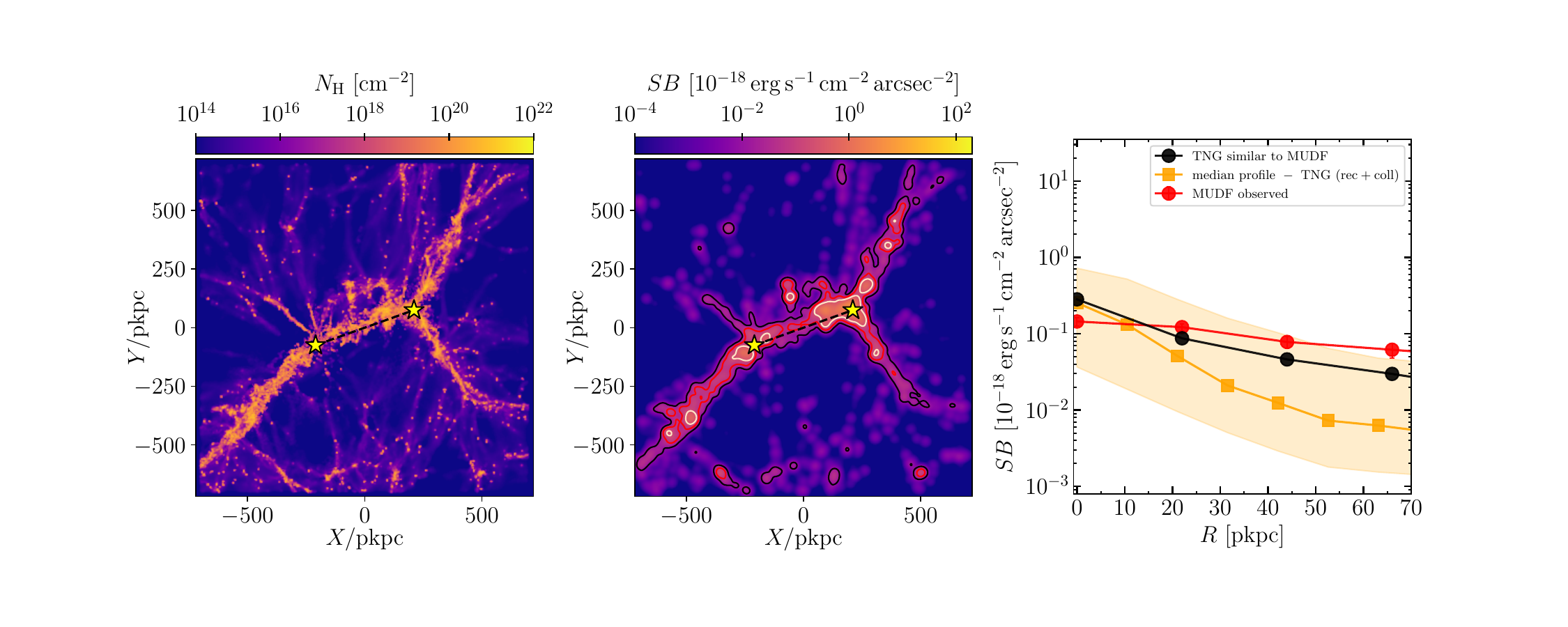}
\caption{\textbf{Simulated pair similar to the one in MUDF}. Example of a simulated pair with a 3D physical distance below 1 pMpc, closely resembling the MUDF system.  \textbf{(a)} The hydrogen column density map, \textbf{(b)} the surface brightness map, smoothed on the same scale as the reconstructed MUDF maps. Contour levels of $10^{-20}$, $10^{-19}$, and $10^{-18}$~\sblcgs\ are marked (black, red, and light-yellow lines). \textbf{(c)} The transverse surface brightness profile of the MUDF twin (black line) compared to the MUDF data (red line). Also shown is the median transverse profile obtained with the pairs with a 3D distance below 1~pMpc (yellow line with shaded regions marking the $16^{\rm th}$ and $84^{\rm th}$ percentiles). The profile is shown up to $\sim 70$ pkpc, where the measurements exceed the detection limit.}
\label{fig:4}
\end{figure}

As two of the main \lya\ emission mechanisms (recombination radiation and collisional excitation) depend quadratically on density, a comparison between the observed and predicted surface brightness of filaments offers a way to constrain the typical order of magnitude underlying gas density within the filaments.
We compute surface brightness maps (see a full description in Methods) under the approximation where recombinations and collisional excitations give the total emissivity of the diffuse gas, i.e., below the densities at which the gas in the adopted model lies on the imposed equation of state ($n_H<0.1~\rm cm^{-3}$). 
We do not include scattering in this baseline model, but note that radiative transfer calculations imply a boost factor of $\approx 2-3$ \cite{Byrohl&Nelson2023} at the observed surface brightness levels. Hence, our main conclusions about the inferred density are not significantly affected.
Due to the presence of the quasars and to test their possible effects, we also consider a maximal fluorescence model, in which bright sources fully ionize the gas that emits only through recombination. We find that at the typical densities of the filaments, the gas is already substantially ionized, and the maximal fluorescence model does not differ significantly from the baseline calculation (see Methods).
Finally, we also study the resolution effects by repeating the same analysis in other boxes of the TNG suite (TNG50-1 and TNG300-1) for a factor of $\sim 200$ in resolution (see Methods).
The predicted surface brightness is generally insensitive to the resolution adopted for the simulations. While large clumping factors (up to $\sim 1000$ \cite{Cantalupo2014}) are often invoked to reproduce the high surface brightness of the quasar nebulae, such as the Slug Nebula, in the low-density regime of the MUDF filament, clumps do not appear essential to reach the observed surface brightness levels. A low filling factor of optically thick clouds within filaments is also in line with the statistics inferred from the absorption lines and the prediction of simulations of the Ly$\alpha$ forest.

Considering our baseline model and following the same methodology adopted in the MUDF analysis, we calculate the median transverse surface brightness profile of filaments between pairs (Extended Data Figure \ref{fig:ext7}).
The observed and simulated profiles share a different normalization. However, raising the filaments' density by less than a factor of three would be enough to match the observed profile, considering the contribution of recombinations alone. A further contribution from collisions reduces this discrepancy, which could even be removed if $n_H>0.1~\rm cm^{-3}$ gas is considered (see Methods), or allowing for a moderate boost from scattering.
Finally, owing to intrinsic differences in the various simulated profiles, we can identify close matches in terms of the transverse surface brightness profile of the MUDF filament inside this simulation. An example of this is shown in Fig. \ref{fig:4}. 

From this comparison, we conclude that the typical density inferred from the cosmic web in these simulations must be of the order of what is found in the MUDF filament, and, at present, there are no obvious indications of discrepancies between observations and the predictions of the cosmic web in the adopted cold dark matter model. Our study, which has offered quantitative measurements of the structural properties of the cosmic filaments at $z\sim3$ beyond a simple detection, exemplifies a tantalizing new direction for constraining the cosmic web with quantitative data to deepen our understanding of one of the most fundamental predictions of the cold dark matter model. 

The remarkable depth, with a detection limit in the deepest part of the narrow-band image of $4.5 \times 10^{-20}$~\sblcgs\, reached by the MUDF observations has enabled the detection of a prominent cosmic filament that connects two halos hosting quasars at $z\approx 3.22$. These observations facilitate a high-definition view of the cosmic web, allowing us to characterize the filament morphology and directly measure the transition radius between the IGM and the CGM, which occurs around the virial radius for $\approx 2-8\times 10^{12}~$M$_\odot$ halos. We also derived the surface brightness profile along the filament and in the transverse direction. With the aid of SAM and cosmological hydrodynamical simulations, we have shown how the MUDF field is an excellent laboratory for studying the physics of general filaments around $\gtrsim 10^{12}-10^{13}~$M$_\odot$ halos at $z\sim3$. Our analysis reveals that a main filament, with gas overdensities above ten times the mean cosmic density, connects most halos with physical separation $<1$~pMpc. This differs from what is seen in pairs at larger separations, e.g., $>2$~pMpc, where the gas between halos drops to the mean density. 

By exploiting the quadratic dependence of the \lya\ emissivity for two main channels of photon production (recombination and collisional excitation), we have used the observed surface brightness maps to test the predicted density distribution of cosmic filaments in the current cold dark matter model. We found a shift between the simulated and observed \lya\ surface brightness levels. However, this difference can be easily removed by raising the underlying density of the filaments by less than a factor of three, demonstrating that the typical densities in models are within acceptable values.
Moreover, we identified examples inside the simulation that closely match the observed MUDF system. Therefore, the current data do not highlight significant tensions with the cold dark matter model. 

By moving from detections to quantitative analysis of the cosmic web, our study demonstrates the exciting potential of spectrophotometry of cosmic filaments for testing how cosmic structures assemble. As the cosmic web is a fundamental prediction of the current cosmological model, a quantitative characterization of its structure and physical properties should be explored more as a way to test the nature of dark matter. Building on our work, future ultradeep observations of cosmic filaments in the era of 40m telescopes coupled with sophisticated numerical models will strengthen our understanding of the Universe in novel ways. 
  

\renewcommand{\figurename}{Extended Data Fig.}
\renewcommand{\thefigure}{\arabic{figure}}   
\setcounter{figure}{0}    

\section*{Methods}\label{methods}

\subsection*{Observations and data reduction}\label{sm:observations}

Observations of the MUSE Ultra Deep Field ($\mathrm{R.A.} =21^{h}$:42$^{m}$:24$^{s}$, $\mathrm{Dec.} =-$44$^{\circ}$:19$^{m}$:48$^{s}$) have been obtained as part of the ESO Large Programme (PID 1100.A$-$0528; PI Fumagalli) between periods 99-109, using the MUSE instrument in wide-field mode with extended wavelength coverage between $\lambda 4650-9300$~\AA. A total of 
358 individual exposures have been collected. Each exposure is dithered around the nominal pointing centers, positioned on the line connecting the two QSOs, and rotated in steps of 5 degrees. The last 60 exposures are centered at six positions surrounding the QSOs, to slightly extend the footprint of the final mosaic while collecting more depth in the central region. This approach, combined with advanced illumination correction algorithms \cite{Cantalupo2019,Fossati2019,Lofthouse2019}, reduces the instrumental signatures in the final co-add produced by the different response of the 24 MUSE spectrographs. Each exposure has an integration time of 1450s except for the first 19 that have been integrated for 1200s, leading to a total observing time of 142.8~h on-source. The final mosaic covers an $\approx 1.5\times 1.5$~arcmin$^2$ area, with maximal sensitivity in the inner $\approx 1$~arcmin$^2$ region (see Fig. \ref{fig:1}). Using the GALACSI adaptive optics system improves the image quality compared to natural seeing, 
yielding a full width at half-maximum of $\approx 0.73$~arcsec for point sources.  

The reduction of MUSE data follows the steps described in the MAGG survey \cite{Lofthouse2019} and articles in the MUDF series \cite{Fossati2019}. Using standard techniques from the MUSE pipeline \citep[v2.8][]{Weilbacher2014}, we corrected the raw data for primary calibrations (bias, dark, flat correction, and wavelength and flux calibration). Next, individual exposures, sky-subtracted and corrected for residual illumination using the {\sc CubExtractor} toolkit (v1.8, {\sc CubEx} hereafter, see \cite{Cantalupo2019} for further details), are coadded, without weighting, in a final datacube with a pixel size of $0.2$~arcsec and $1.25$~\AA\ in the spatial and spectral direction, respectively. Prior to the final coaddition, the pixels at the edges of the slitlets, which are affected by slight vignetting, are masked and each exposure is inspected. In two exposures, satellite trails have been encountered and masked.
An associated variance cube is also reconstructed using the bootstrap technique developed in previous work \cite{Lofthouse2019,Fossati2019}. The final data achieve a depth of up to $110$ h and a pixel root-mean-square (rms) of $3\times10^{-21}$~\fcgs~pix$^{-1}$ (at $\sim 5200$~\AA, see below), making this observation comparable to the MUSE eXtremely Deep Field \cite{Bacon2021}.

Due to the exquisite sensitivity of these data and the fact that we are interested in extended low surface brightness emission, we minimize the impact of small sky residuals by performing a final correction of the background level. For this, we select pixels in regions empty of continuum sources and far from where we expect \lya\ emission. After identifying the extended emitting structure in the datacube (see next section), we explicitly check that no source contribution is contained in these sky regions. Next, we construct a median sky spectrum over $\approx 5000$ pixels and use this spectral template for the final sky subtraction. Locally, this template is normalized to the residual sky values measured in two narrow-band (NB) images of spectral width $60$~\AA, which we select around $\lambda=4800$~\AA, and $\lambda=5400$~\AA, far enough from the wavelength interval where we expect \lya\ emission at $z\approx 3.22$. Ultimately, we achieve robust sky subtraction, where the residual level is $<1\%$ of the pixel rms measured in the deepest central region of the field of view ($>95~\rm h$) and in the spectral range 4900-5100~\AA\ and 5200-5400~\AA\ that is adjacent to the wavelengths where we expect \lya\ emission. 

\subsection*{Identification and extraction of the filament and nebulae}\label{sm:detection}

The presence of two bright \lya\ emitting nebulae around the two MUDF quasars was already confirmed in a partial, $\approx 40$~hour, dataset by Lusso et al. \cite{Lusso2019}. To search for more extended and very low-surface brightness emission in the datacube, we use the {\sc CubEx} tool to subtract continuum-emission sources and the quasar point spread function through a nonparametric continuum-subtraction algorithm (see \cite{Cantalupo2019,ArrigoniBattaia2019} for further details).
Next, we identify groups of $>2500$ connected voxels above a signal-to-noise ($S/N$) threshold of 2, with a minimum number of $500$ spatial pixels. To increase the sensitivity to low-surface brightness emission, we further smooth the cube with a Gaussian kernel with a size of $3$ pixels ($0.6$ arcsec or $4.6$ pkpc) in the spatial direction, masking continuum sources to avoid contamination from negative or positive residuals created during the continuum subtraction process. No smoothing is applied in the wavelength direction to maximize the spectral resolution. We also demand that at least $3$ wavelength layers be connected in the spectral direction to avoid spurious thin sheets of emission. 

This search yields two connected and very extended structures ($>30000$ voxels). One of them covers the two quasars, with the brightest emission coinciding with the quasar positions. The other is the previously known Nebula 3 at a different redshift of $3.254$ by Lusso et al. \cite{Lusso2019}. 
The quasars \lya\ nebulae are detected to a surface brightness limit of $\sim 1\times10^{-19}$~\sblcgs~ in the new data. With ultradeep observations in the central region, we also uncover a filamentary and extended emission signal that originates from the edges of the quasar nebulae and connects them in the direction of the putative filament proposed by Lusso et al. \cite{Lusso2019}. Additional extended emission is detected in the opposite direction for both quasars, suggesting that the emitting structure extends for more considerable distances than probed by our data.
The detection of the nebulae and filaments with their overall morphology does not depend on the selection criteria described above. In fact, we perform the extraction using different $S/N$ thresholds (up to 2.5) and different spatial (2 and 4 pixels) and spectral (0 and 1 pixel) smoothing settings, all of which yield the same global emission structures once the different parameters employed are considered.

\begin{figure}
  \begin{minipage}{1\textwidth}
    \centering
    \begin{subfigure}{0.9\textwidth}
        \includegraphics[width=\textwidth]{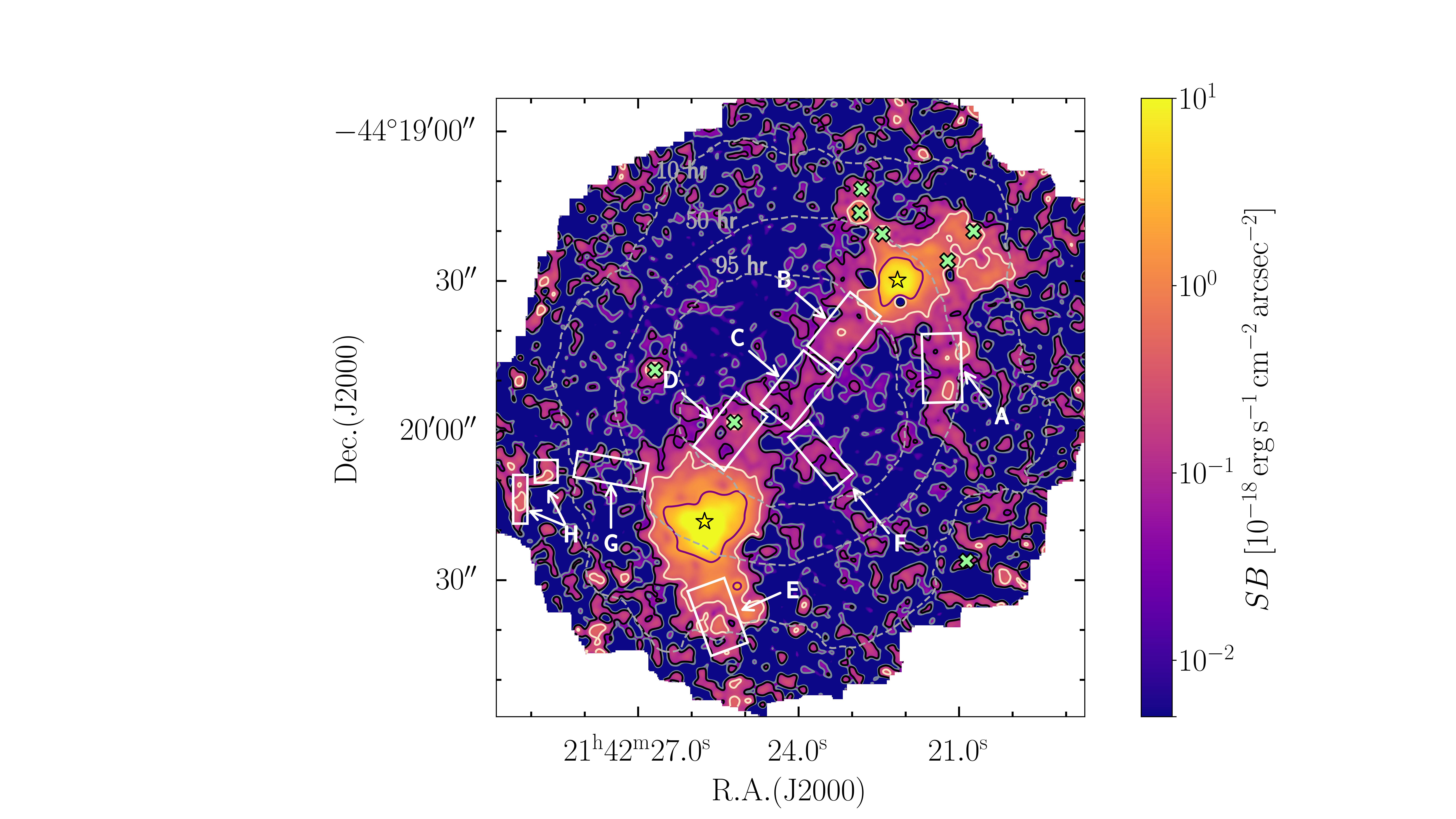}
        \caption{}
        \label{fig:ext1a}
    \end{subfigure}
  \end{minipage}
  \begin{minipage}{1\textwidth}
    \centering
    \begin{subfigure}{1\textwidth}
        \includegraphics[width=\textwidth]{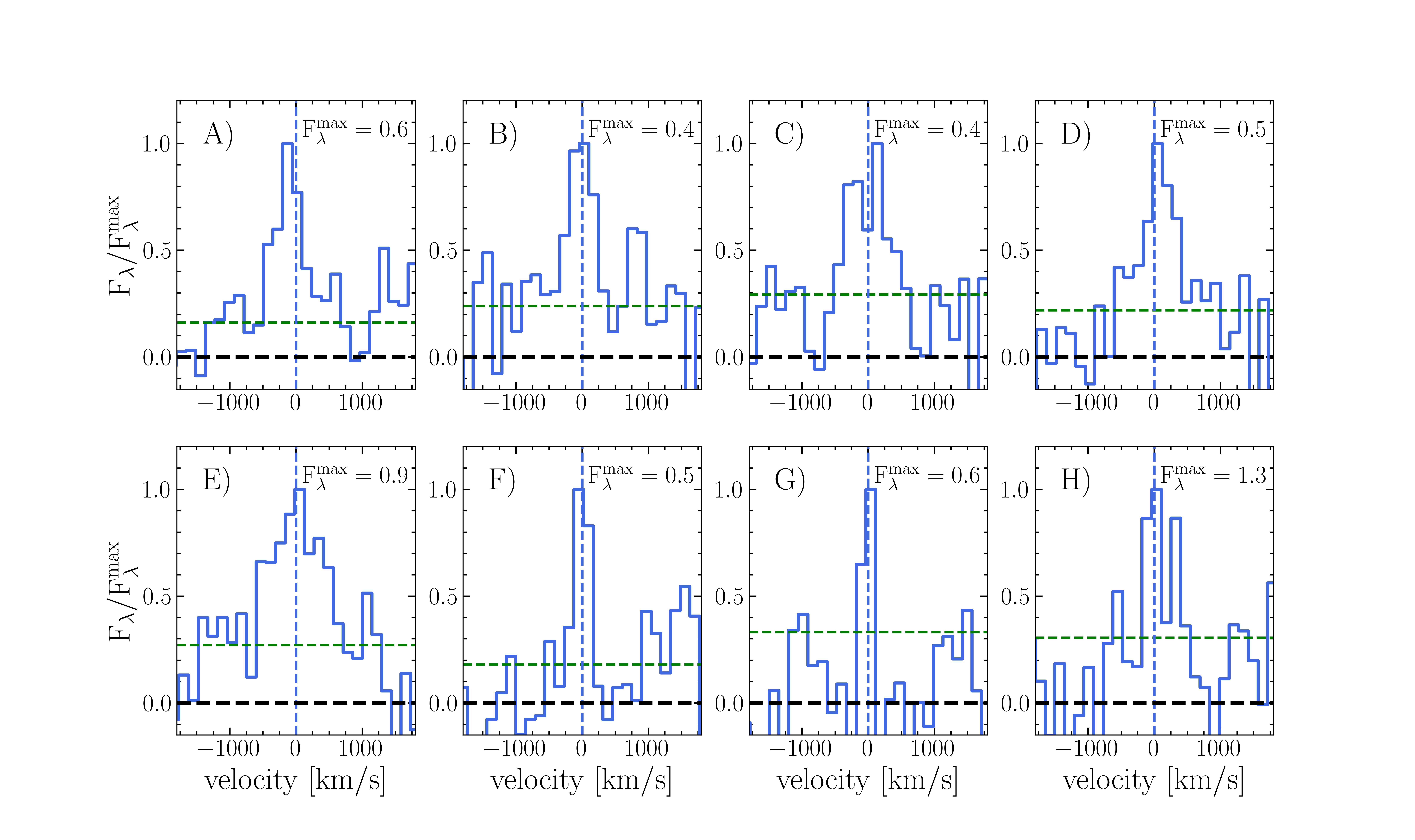}
        \caption{}
        \label{fig:ext1b}
    \end{subfigure}
  \end{minipage}
  \caption{\textbf{\lya~emission image with noise and mean spectra}. \textbf{(a)} Optimally extracted \lya\ image on top of three collapsed wavelength layers at the central wavelength of the quasar nebula emission. The contour levels are $0.02, 0.06, 0.3$ and $2 \times 10^{-18}$~\sblcgs\ (light-grey, black, light-orange, purple). The green crosses represent the position of the detected \lya\ emitters within $1500~\rm km~s^{-1}$ from the average redshift of the two quasars. \textbf{(b)} Normalized mean spectra in velocity space extracted from the white boxes labeled from A to H in (a). The $F^\mathrm{max}_\lambda$ are the maximum flux density values used to normalize the mean spectra in units of $10^{-21}$ erg s$^{-1}$ cm$^{-2}$ \AA$^{-1}$. The vertical dashed blue line represents the zero velocity reference calculated from the first moment of the line, and the horizontal dashed green line represents the $1\sigma$ noise level of the spectrum.}
  \label{fig:ext1}
\end{figure}

The detected signal is then projected along the wavelength direction to compose an optimally extracted \lya\ image, shown at the top of the white-light image in Fig. \ref{fig:1}, and at the top of three collapsed wavelength layers at the central wavelengths of the nebulae to assess the background noise level in Extended Data Figure \ref{fig:ext1}. Optimally-extracted maps are best suited to highlight low surface brightness emissions. Still, since they combine only the voxels identified by {\sc CubEx} above the $S/N$ threshold as harboring significant emission, they may lose flux and are inadequate for precise estimates of the total surface brightness (see, e.g., \cite{Borisova2016}). Therefore, we resort to a synthesized NB image centered on the wavelength at the redshift of the \lya\ emission peak and obtained by summing the flux in the wavelength direction over $30$~\AA\ for all the measurements presented in this work.
Hereafter, we will use the term Nebula 1 (Nebula 2) to refer to the one associated with the brighter (fainter) quasar, QSO1 (QSO2), as indicated in Fig. \ref{fig:1}.

While the nebulae are detected at high $S/N$, the extended low-surface brightness signal across the filament, with $\approx 8 \times 10^{-20}$~\sblcgs, is close to the detection limit of the NB image ($\approx 2\sigma$, see below in the section \nameref{sm:analysis} how the detection limit $\sigma$ is defined in the NB image). Therefore, we perform a series of tests to confirm the genuine nature of the detection. Firstly, we search again for the detected signal in a primary data reduction before applying any illumination correction or enhanced sky subtraction using the {\sc CubEx} code. Secondly, we verify the presence of the extended structure in two independent coadds containing each half of the total number of exposures.  The filament is recovered in each test, although at lower $S/N$ due to the higher noise of the various products and methods used for this test. 
Finally, we verify that the signal along the main filamentary structure connecting the two quasars represents a significant spectral feature in the datacube. By masking pixels associated with continuum sources, we extract the mean spectra from five distinct regions, labeled A, B, C, D, and E, each measuring approximately $14 \times 8$ arcsec$^2$, and positioned along the main filament, extending from the vicinity of QSO2 up to QSO1, as illustrated in Extended Data Figure \ref{fig:ext1a}. 
Additionally, we consider three other apertures, labelled as F and G, each $\approx 20 \times 8$ arcsec$^2$ and H composed of two apertures of $\approx 7 \times 7$ arcsec$^2$ and $\approx 15 \times 4$ arcsec$^2$, respectively, positioned along the thin protuberances branching from the main filament, as illustrated in Extended Data Figure \ref{fig:ext1a}.

In Extended Data Figure \ref{fig:ext1b}, we present the extracted normalized spectra for each extraction box. The spectra are shown in velocity space, with the reference zero velocity (marked by the vertical dashed blue line) calculated from the first moment of the line within a wavelength range $\pm 10$~\AA\ around the \lya~ emission peak.  
The dashed horizontal green line represents the 1$\sigma$ noise level of the spectrum estimated from the wavelengths not in the interval of the \lya\ emission. The spectra extracted in the regions along the main filamentary structure confirm that the detected signal is an actual emission of astrophysical origin and does not arise from spurious noise or systematic artifacts in the advanced processing of the data. Additionally, we spectroscopically confirm the presence of the protuberance in box F.  Box G, including pixels connected to QSO1 by the algorithm, does not show a clear emission line at the present depth, but aperture H reveals a detectable signal, though it is more affected by noise. 

Using these spectra, we also assess the kinematic properties of the filament.
We examine the first moments of \lya\ emission lines extracted along the main filamentary structure using boxes A, B, C, D, and E, comparing them to the wavelength at the redshift of the \lya\ emission peak. No significant velocity gradient along the main filamentary emission structure is detectable, within the errors and at the present depth ($\lvert \Delta v \rvert < 100$ km/s). In Extended Data Figure \ref{fig:ext1a} we also mark the positions of the detected and spectroscopically confirmed Lyman-alpha emitters (LAEs) within $1500$ km/s of the average systemic redshift of the two quasars with green crosses. In the main filamentary structure connecting the two quasars, where our analysis is focused, we identify only one LAE. Whenever we mask continuum sources, including during the spectra extraction, we also exclude compact emission from this LAE.

\subsection*{Analysis of the surface brightness profiles}\label{sm:analysis}

\begin{figure}
  \begin{minipage}{0.5\textwidth}
    \centering
    \begin{subfigure}{1\textwidth}
        \includegraphics[width=\textwidth]{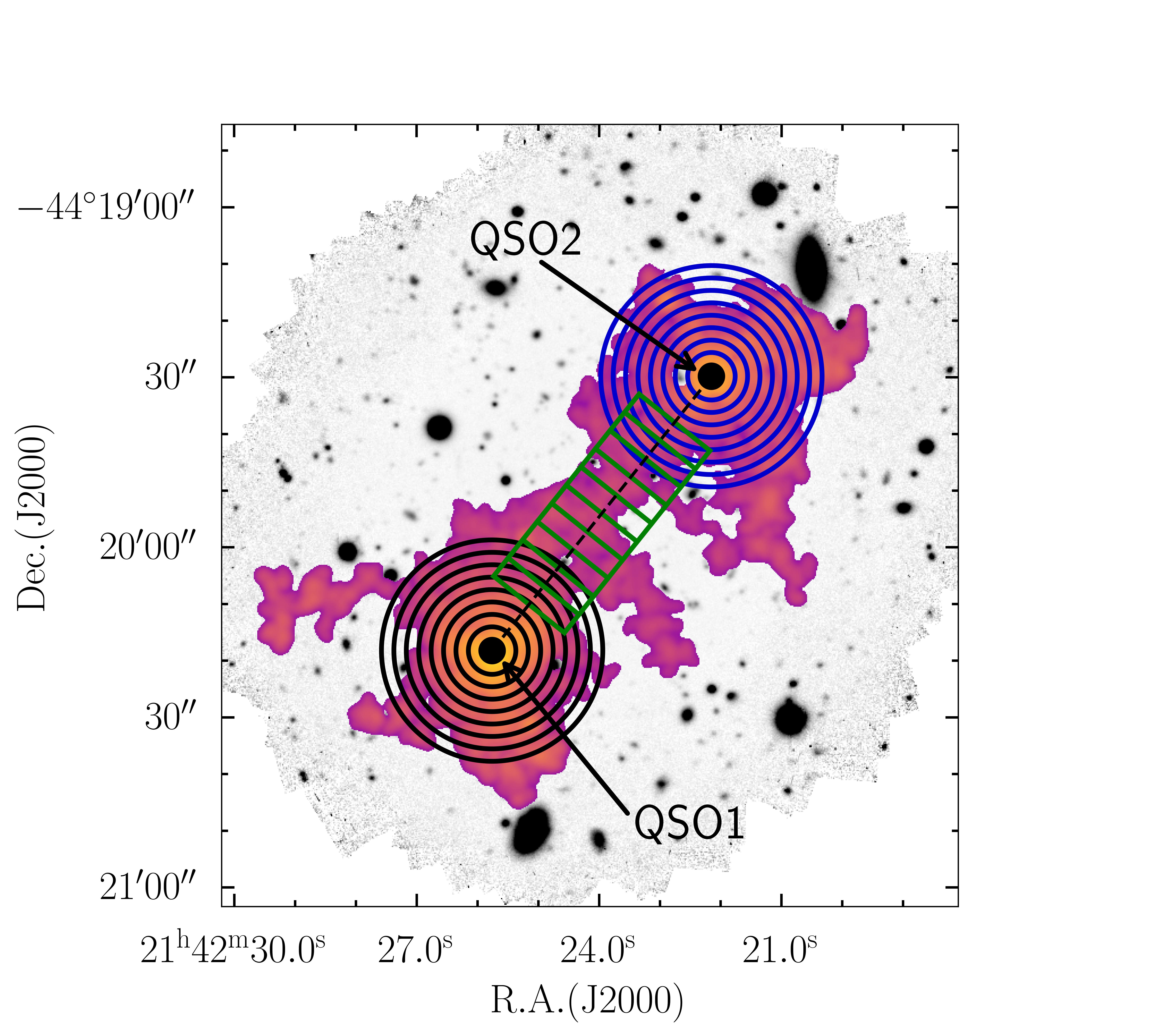}
        \caption{}
        \label{fig:ext2a}
    \end{subfigure}
  \end{minipage}\hfill
  \begin{minipage}{0.5\textwidth}
    \centering
    \begin{subfigure}{1\textwidth}
        \includegraphics[width=\textwidth]{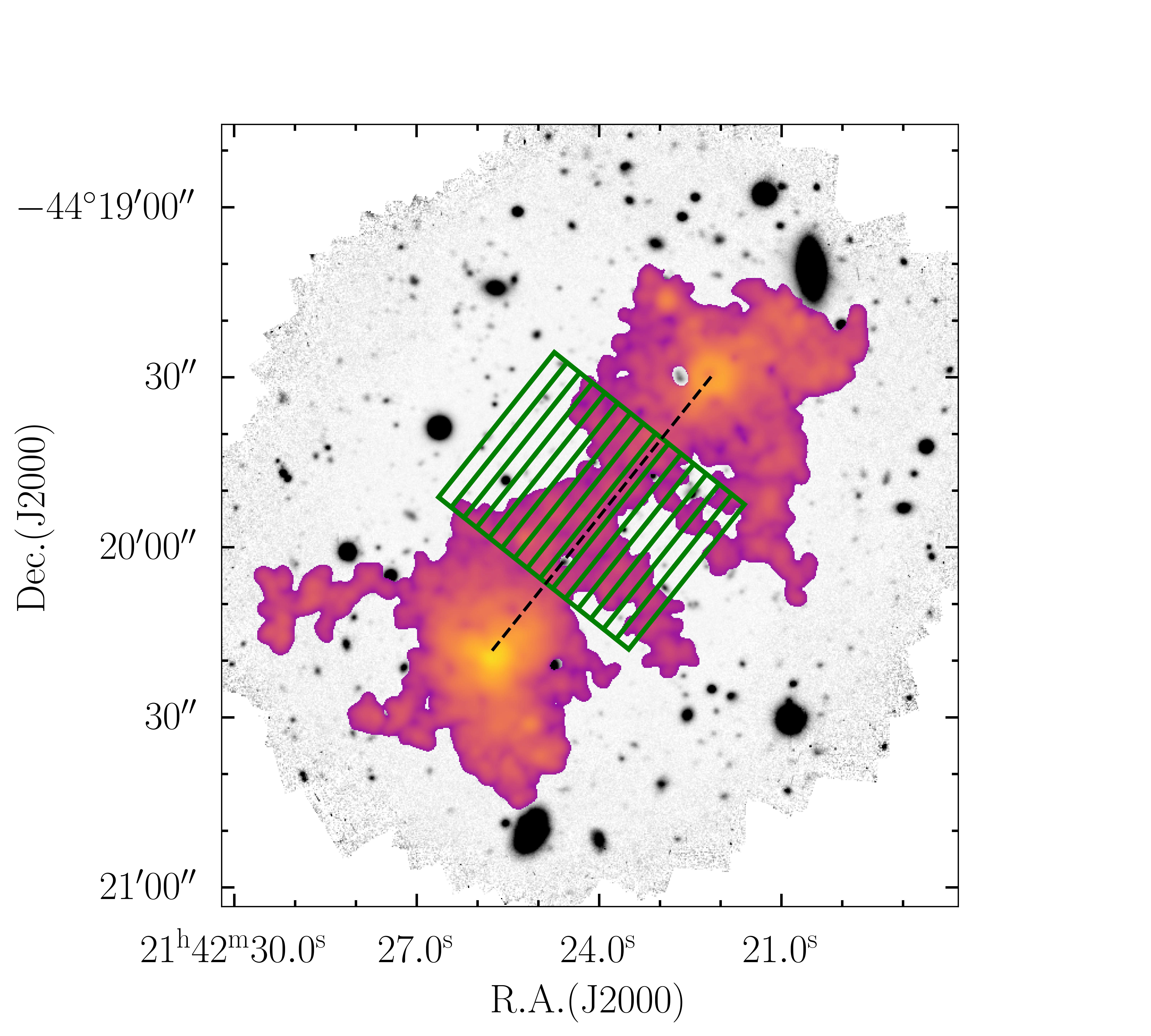}
        \caption{}
        \label{fig:ext2b}
    \end{subfigure}
  \end{minipage}
\caption{\textbf{Extraction apertures for surface brightness profiles of nebulae and filament.} \textbf{(a)} Extraction apertures (black annuli for Nebula 1, blue annuli for Nebula 2, and green boxes for the filament) used to extract the surface brightness profiles, superimposed to the \lya\ emission map, included solely for illustrative purposes. The two black dots mark a $\approx 15$~pkpc radius where the quasar's PSF residuals dominate the signal. QSO1 and QSO2 are indicated by the two black arrows, respectively. \textbf{(b)} Same as (a) but for the transverse surface brightness profile. In the background of both images, shown in grey, is a white-light image of the region imaged by MUSE.}
\label{fig:ext2}
\end{figure}

We compute the surface brightness profiles of the nebulae associated with the two quasars and the filament from the synthesized NB image. Before extracting the surface brightness profiles, we apply a mask to avoid residuals of the continuum subtraction of compact sources, which could introduce contamination. The percentage of unmasked pixels in each region used for the analysis is $>80$~percent.
Subsequently, for each nebula, we compute the circularly averaged surface brightness profile, using rings with an aperture of $\approx 10$ spatial pixels (equivalent to $\approx 2$~arcsec, or $\approx 15$~pkpc), up to a radius of 150~pkpc centered at the positions of the quasars. In Extended Data Figure \ref{fig:ext2a}, solely for illustrative purposes, we overplot the annuli on the optimally extracted image within which we calculate the surface brightness profiles (in black for Nebula 1 and blue for Nebula 2). We exclude a circular $2$~arcsec$^{2}$ region around the quasars to avoid residuals from the point spread function subtraction (the two black dots). Hence, the radial profiles are presented for $R \gtrsim 15$~pkpc. Instead, we use ten boxes for the filament, each with dimensions of approximately $20 \times 80$ pixels$^2$ along the direction connecting the two quasars, starting at a distance of $80$~pkpc from them. These boxes are shown in green in Extended Data Figure \ref{fig:ext2b}.

To calculate the surface brightness value for each annulus and box, we adopt a variance-weighted average of all unmasked pixels, consistently propagating the errors from the associated variance image. 
Detection limits are computed radially, taking into account the effective exposure time and the number of pixels in the regions considered for this analysis. We also explicitly verify that no residual background signal is detected in empty regions above the detection limit. 
Considering the varying aperture widths and shapes used and the irregular geometry of the emission, we examine the overlapping regions between the rings and the boxes to join the nebula and the filament profile. 
In Fig. \ref{fig:2a}, the observed surface brightness data points are shown using the same colors as for the extraction apertures in Extended Data Figure \ref{fig:2a}: black dots represent the measured values in the annuli for Nebula 1, blue dots represent those for Nebula 2, and green dots represent those for the boxes in the filament. The filled dots mark the combined surface brightness profiles, while the empty dots are the individual aperture measurements not included in the final profile and up to the detection limit of the NB image.

The profile of both quasars is rapidly decreasing to radii of $\approx 100$~pkpc, which we identify as the \lya\ nebulae arising from the circumgalactic medium (CGM) of the host. The plateau at $\gtrsim 100$~pkpc extending to $250$~pkpc -- the midpoint between the two quasars -- instead arises from the filament. Examining the empty data points of the two nebulae above $\sim 100$~pkpc, the excess emission from the filament becomes evident compared to what is measured in the annuli. Clearly, the filament signal is also enclosed in the annuli, but the filling factor becomes progressively low as the radius increases, making this circular geometry a poor choice for the filament surface brightness. Moreover, the annuli include the signal on the opposite side of the nebulae, where the filamentary structure is fainter, yielding a steeper profile. 

In Fig. \ref{fig:2b}, we compare the observed profiles of the two nebulae on a log-log scale with the average profile from the $z\sim3.3$ sample of Borisova et al. \cite{Borisova2016} (dashed magenta line), the $z\sim3.2$ sample of Arrigoni Battaia et al. \cite{ArrigoniBattaia2019} (dashed blue line) and the $z\sim3.16$ faint sample of Mackenzie et al. \cite{Mackenzie2021} (dashed sky blue line). We observe that both our nebulae have a profile that is in excellent agreement with those in the literature at the same redshift, once accounting for the different magnitudes of each quasar. Infact, the $i-$band magnitudes of the quasars in the sample from Borisova et al. \cite{Borisova2016} range from $16.6$~mag to $18.6$~mag. Those of Arrigoni Battaia et al. \cite{ArrigoniBattaia2019} range from $17.4$~mag to $19$~mag and include in the sample the brighter QSO1, with a $i-$band magnitude of $17.6$~mag. Finally, the $i-$band magnitude of the quasars in Mackenzie et al. \cite{Mackenzie2021} ranges from $20$~mag to $23$~mag, thus being comparable to the fainter QSO2, with a magnitude of $20.6$~mag in the same band. We conclude that both quasars have a typical CGM when traced by \lya\ despite being in a close pair. 

An evident inflection point is visible in both profiles around $\approx 100$~pkpc from the quasars, which we ascribe to the transition between the CGM traced by the quasar's nebulae and the IGM traced by the filament. To explore this transition more and to separate the emission of these two components, we fit the complete surface brightness profiles (filled dots) with the following broken power law model
\begin{equation}\label{eq:pwl}
    SB(R;A,R_\mathrm{t},b_1, b_2) = \begin{cases} 
    A \left( \frac{R}{R_\mathrm{t}} \right)^{b_1}, & \text{if } R \leq R_\mathrm{t}, \\ 
    A \left( \frac{R}{R_\mathrm{t}} \right)^{b_2}, & \text{if } R > R_\mathrm{t}. \end{cases} 
\end{equation}
 $R_\mathrm{t}$ is the transition radius between the CGM and the IGM, $A$ is the normalization, and $b_1$ and $b_2$ are the two slopes. The choice of a power law function is justified in the quasar \lya\ nebulae literature context (e.g., \cite{Borisova2016}, \cite{ArrigoniBattaia2019}).
To determine the best parameters, we employ a Bayesian approach, assuming a Gaussian likelihood for each surface brightness estimate and a uniform prior.
The best-fitting parameters obtained through the \textit{emcee} algorithm (\cite{Foreman-Mackey2013}) are reported in Table~\ref{tab:best_fit_params}.

\begin{table}[h!]
\centering
\begin{tabular}{lcc}
\hline
Parameter & QSO1 & QSO2 \\
\hline
$R_\mathrm{t}$ (pkpc) & $117 \pm 8$ & $108 \pm 10$ \\
$A$ ($10^{-18}$ \sblcgs) & $0.18 \pm 0.04$ & $0.09 \pm 0.03$ \\
$b_1$ & $-3.3 \pm 0.1$ & $-2.71 \pm 0.03$ \\
$b_2$ & $-0.9 \pm 0.3$ & $0.1 \pm 0.4$ \\
\hline
\end{tabular}
\caption{\textbf{Best-fitting parameters for the surface brightness profiles of QSO1 and QSO2.} Data are reported as median values, along with the $16^{\rm th}$ and $84^{\rm th}$ percentiles.}
\label{tab:best_fit_params}
\end{table}

The best fits are shown in Fig. \ref{fig:2a} as black and blue dashed lines for QSO1 and QSO2. The solid vertical lines, with the same colors, represent the transition radius and its error. Instead, the dotted black and blue lines represent the extrapolation of the best fit for the nebulae and the filament for each quasar, further confirming the presence of an emission excess in the filament region.
This model thus recovers a natural transition between the CGM and the IGM in the range of $\approx 90-130$~kpc, with the fainter quasar having a smaller size of the \lya-emitting CGM as observed in previous studies (\cite{Mackenzie2021,ArrigoniBattaia2023}).
For typical halo masses in the order of $\approx 10^{12.3-12.5}$~\msun\ for these quasars (\cite{Fossati2021,deBeer2023}), the virial radius at $z\approx 3.2$ is $R_{\rm vir}=92-108$~kpc, i.e., comparable with the size of the transition radius. The transition occurs at a surface brightness of $\approx (1-2)\times 10^{-19}$~\sblcgs, with only a small difference with the quasar luminosity. 

In Fig. \ref{fig:2b} we also compare the observed SB profile with stacked profiles of LAEs (dotted lines) obtained in \cite{Kikuchihara2022} and in \cite{LujanNiemeyer2022}. A similar flattening at large radii is observed in these analyses, although at different radii ($40-60$ kpc). The mean \lya\ luminosity of the LAEs studied by Kikuchinara et al. \cite{Kikuchihara2022} at $z=3.3$ is $\log(L_\mathrm{Ly\alpha}/\mathrm{erg\,s^{-1}}) = 42.5$ and the sample in Lujan-Niemeyer et al. \cite{LujanNiemeyer2022} is divided in low ($\log(L_\mathrm{Ly\alpha}/\mathrm{erg\,s^{-1}}) < 43$) and high ($\log(L_\mathrm{Ly\alpha}/\mathrm{erg\,s^{-1}}) > 43$) luminosity, with a median $z=2.5$. After rescaling by surface brightness dimming \cite{LujanNiemeyer2022}, we observe that the SB profiles in Kikuchinara et al. and the sample in Lujan-Niemeyer et al. lie below the observed MUDF profile by a factor of $3-4$ at high luminosity and up to a factor of $10$ at low luminosity. Moreover, we observe that QSOs have a higher emission profile, especially in the inner regions (i.e., their CGM) than LAEs. This discrepancy can originate from the different halo masses probed ($\sim 10^{12.5}~M_\odot$ for quasars and $10^{10-11}~M_\odot$ for LAEs) as well as different ionizing fields. A further explanation for the observed discrepancy at larger radii can derive from the stacking technique used, different from our direct measurement of a single structure connecting two massive halos. Indeed, in stacks, there can be a signal dilution when coadding structures that are not fully aligned.

The depth and quality of the data also allow us to extract the transverse surface brightness profile of the filament. We calculate the weighted average value from boxes measuring $\approx 160 \times 14$~pixels$^2$ each, up to a distance of $165$~pkpc on the right and left sides relative to the direction connecting the two quasars. The 15 boxes are off-axis by 7 pixels in the NE direction to capture the emission peak at $R=0$. To account only for the filament emission and to avoid contamination from the two nebulae, the length of each box is determined by the distance from the two quasars, selected as the transition radius estimated above (see Extended Data Figure \ref{fig:ext2}). The right and left surface brightness profiles are shown in Fig. \ref{fig:3} with gray dashed and dotted lines, respectively. A solid green line shows the combined average profile. Fitting the profile in the region above the detection limit with a power law, we obtain a slope of $-0.74 \pm 0.15$, and we observe that the transverse projected width of the filament extends up to approximately $70$~pkpc without reaching a clear edge at the depths of our observations.

As the formal error does not fully account for the pixel covariance arising from the cube reconstructions \cite{Fossati2019}, we calculate an empirical surface brightness detection limit profile that considers the different extraction apertures and the varying mean exposure times within them. To achieve this, we extract an NB image of 30~\AA, shifted by approximately 60~\AA\ from the \lya\ peak wavelength emission. This spectral region is chosen because we do not expect any source emission.
First, to determine the SB limit associated with the extraction box used for the profiles of the main filament, we focus on the central region of the field of view, specifically in the area with the deepest data (exposure time $>95$~h), where the emitting filament is detected. 
We obtain the distribution of the average SB values along 1000 box apertures, masking any residuals of the continuum subtraction and requiring a percentage of unmasked pixels higher than $95\%$. 
The NB detection limit ($1\sigma$) is then defined as the standard deviation of the distribution above the mean value. 
To determine the SB limit associated with the extraction annuli used for the profiles that characterize the two nebulae around the quasars, we apply the same procedure described above, with the additional condition that the average exposure time of each randomly located annulus is within $10$ h of the average exposure time of the reference annulus in the NB around the \lya\ signal, which is $\approx 60-65$~h. The resulting detection limit profile is shown by the sky blue region in Fig. \ref{fig:2} and Fig. \ref{fig:3}. In the deepest region of the field, this detection limit reaches $4.5 \times 10^{-20}$~\sblcgs, ensuring that the measured \lya~ SB profile is always statistically significant.

To measure the global properties of the detected emission, we use the transition radius as a reference to differentiate between the nebulae and the filament. The nebula around QSO1 has a total flux of approximately $1.1 \times 10^{-15}$~\flcgs\ over a total area of approximately $635$~arcsec$^2$, while the nebula around QSO2 has a total flux of approximately $3.8 \times 10^{-16}$~\flcgs\ over a total area of approximately $530$~arcsec$^2$. This leads to a total luminosity of $\sim 1 \times 10^{44}$~erg~s$^{-1}$ for Nebula 1 and $3.6 \times 10^{43}$~erg~s$^{-1}$ for Nebula 2.
The total linear extension of the emitting structures is also calculated considering the projected distance between the two quasars and the projected distance up to $\sim 1 \times 10^{-19}$~\sblcgs~on the opposite side of the filament, leading to $\sim 700$~pkpc.
Table \ref{tab:global_properties} summarizes these global properties. 

The filament, considered up to approximately $70$~pkpc in the transverse direction, has an average surface brightness of approximately $8.3\times 10^{-20}$~\sblcgs, similar to the average levels of $\approx 3.5 - 11 \times 10^{-20}$~\sblcgs\ found by Bacon et al. \cite{Bacon2021} around groups of LAEs at $z\approx 3-4$. 
Given the observed surface brightness level and the presence of two bright quasars, we can exclude the fact that the bulk of the emitting gas is optically thick, as simple scaling relations (see \cite{Hennawi2013,ArrigoniBattaia2016}) would suggest much brighter emission. Instead, the observed signal aligns more closely with an optically thin scenario, a hypothesis we will corroborate next using hydrodynamic simulations.

\subsection*{Analysis of the semi-analytic model}\label{sm:sam}
As the MUDF was selected for the particular configuration of two closely spaced quasars, we aim to understand how common halos traced by the two quasars are in a CDM Universe to assess whether the properties of the MUDF filament can be used to learn about the IGM connecting halos at this mass scale.
Thus, we turn to the analysis of semi-analytical models (SAM) with two objectives: to place our system in a broader cosmological context by assessing the expected number density of MUDF-like pairs and to infer the most probable distributions of dark matter halo masses of the MUDF quasars. With these distributions, we will consider a hydrodynamic simulation to place constraints on the filament gas density. 

For these tasks, we use a lightcone generated with the updated version of the \texttt{L-Galaxies} SAM models \cite{Henriques2015}, as detailed in Izquierdo-Villalba et al. \cite{Izquierdo-Villalba2019}, \cite{Izquierdo-Villalba2020}.
These models are run using the sub-halo merger trees from the \texttt{Millennium} \cite{Springel2005} dark matter N-body simulation within a periodic cube of side $500$~cMpc/$h$. 
The quasar phase of a galaxy is triggered by gas accretion onto black holes, and this model accounts in detail for these processes, reproducing statistics in good agreement with observations, including the literature quasar luminosity functions (see Izquierdo-Villalba et al. \cite{Izquierdo-Villalba2020} for a more comprehensive discussion). Therefore, we can leverage these models to identify systems similar to the MUDF. 

\begin{figure}
  \begin{minipage}{0.5\textwidth}
    \centering
    \begin{subfigure}{1\textwidth}
        \includegraphics[width=\textwidth]{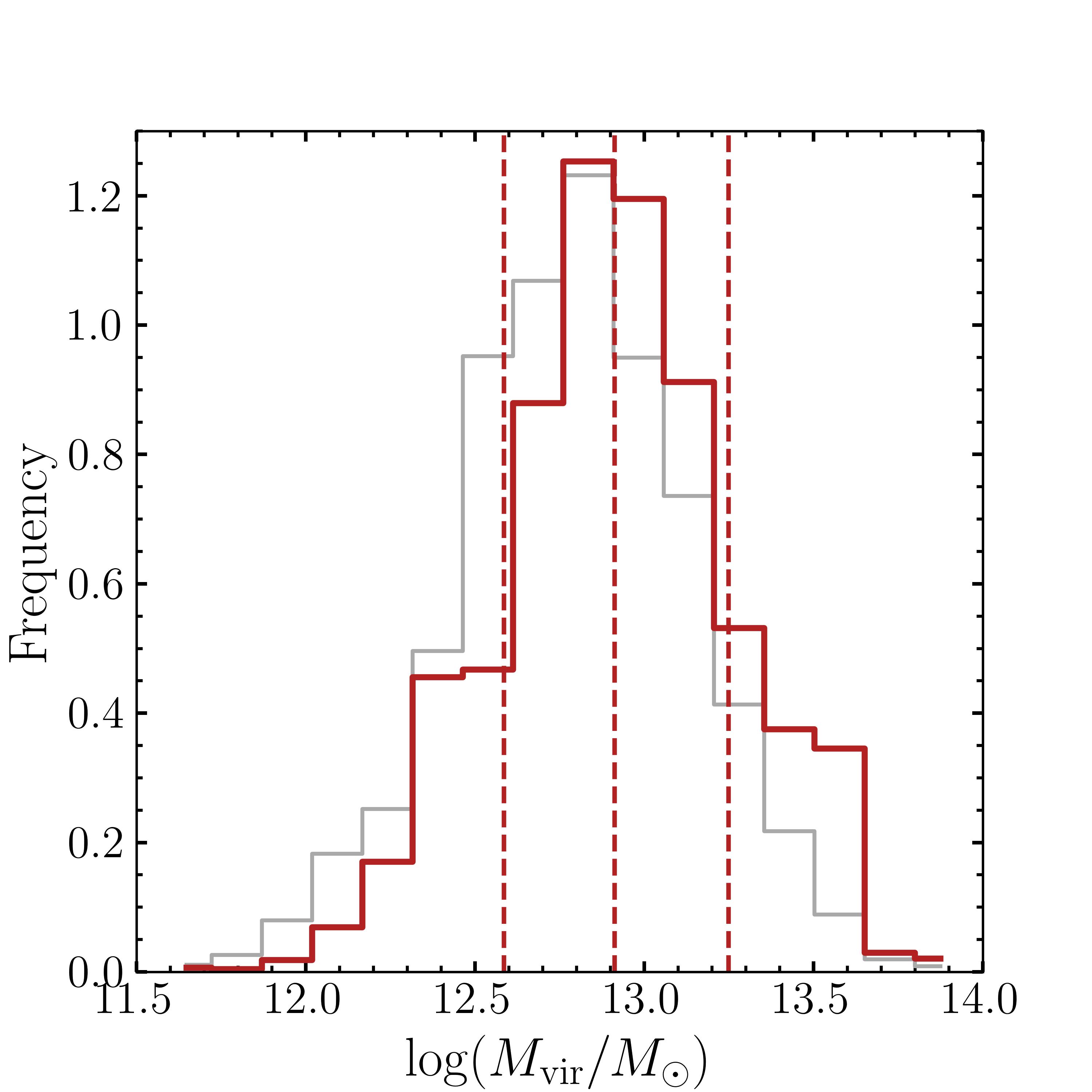}
        \caption{}
        \label{fig:ext4a}
    \end{subfigure}
  \end{minipage}\hfill
  \begin{minipage}{0.5\textwidth}
    \centering
    \begin{subfigure}{1\textwidth}
        \includegraphics[width=\textwidth]{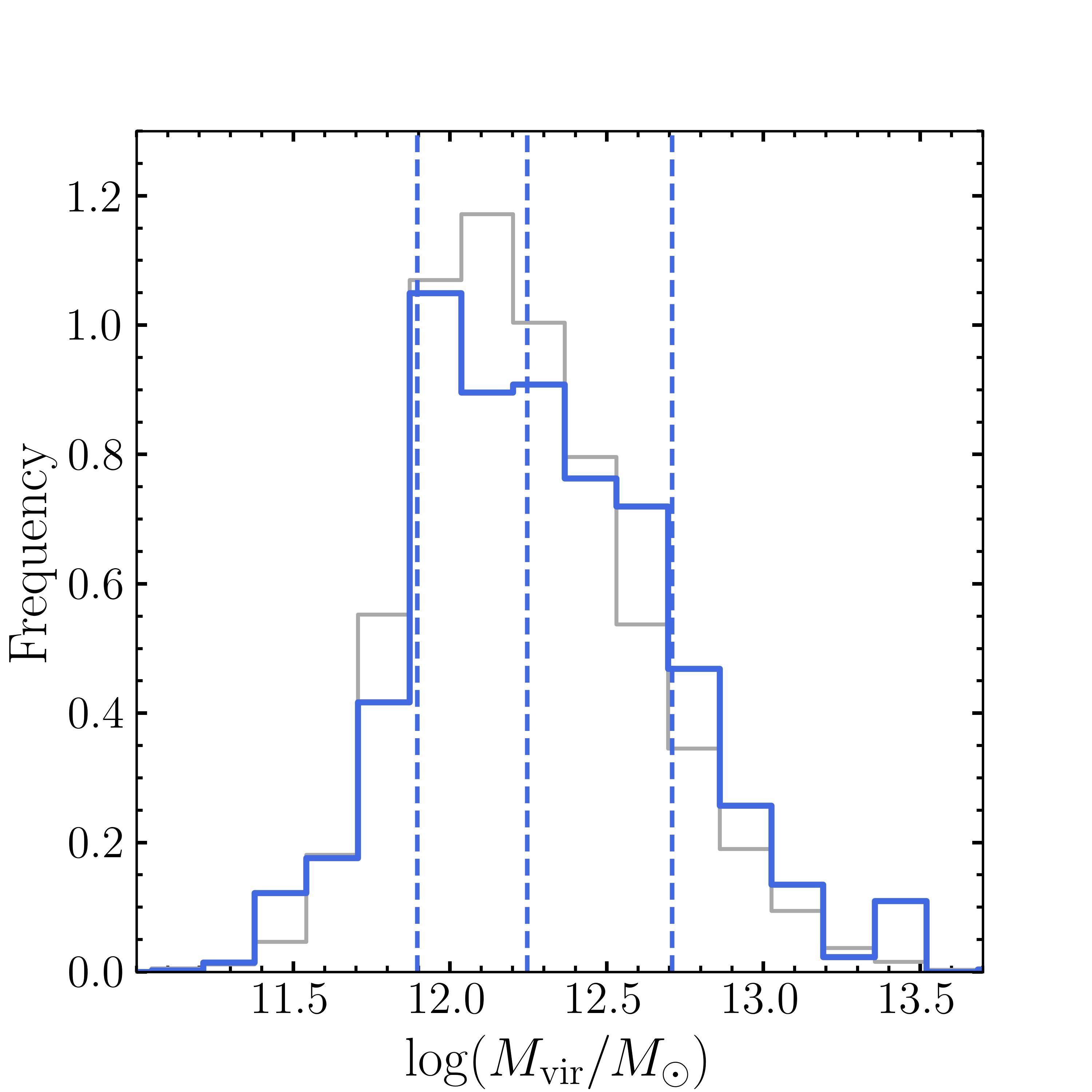}
        \caption{}
        \label{fig:ext4b}
    \end{subfigure}
  \end{minipage}
  \caption{\textbf{Virial halo mass distributions for bright and faint quasars.} \textbf{(a)} The virial halo mass distribution in the SAM of the bright quasars ($\log(L_\mathrm{bol}/{\rm{erg}}\,{{\rm{s}}}^{-1}) = 47.3 \pm 0.3$) selected across the full-sky is shown in grey. The red curve represents the distribution of bright quasars among the selected pairs that mimic the MUDF system in the sky. \textbf{(b)} Same as (a) but for the faint quasars ($\log(L_\mathrm{bol}/{\rm{erg}}\,{{\rm{s}}}^{-1}) = 46.3 \pm 0.3$). In blue, the subsample resembling the MUDF system in the sky. All distributions are normalized to the bin width and the number of bright and faint sources obtained from their respective selections. Dashed lines indicate the mean values along with the $16^{\rm th}$ and $84^{\rm th}$ percentiles.}
  \label{fig:ext4}
\end{figure}

Using the methodology presented by Izquierdo-Villalba et al. \cite{Izquierdo-Villalba2019}, we have created a lightcone covering the full sky in the redshift range $z\sim2.8 - 3.8$, which is centered on the mean redshift of the two MUDF quasars ($z\sim3.22$).
Next, we select the bright quasars corresponding to QSO1 with a bolometric luminosity of $\log(L_\mathrm{bol}/{\rm{erg}}\,{{\rm{s}}}^{-1}) = 47.3 \pm 0.3$ and the faint sources corresponding to QSO2 with luminosity $\log(L_\mathrm{bol}/{\rm{erg}}\,{{\rm{s}}}^{-1}) = 46.3 \pm 0.3$ \cite{Lusso2023}. This results in a number density of approximately $3\times10^{-7}$ cMpc$^{-3}$ and $8\times10^{-6}$ cMpc$^{-3}$, respectively. The virial mass distribution for the two samples has a mean value of $\log(M_\mathrm{vir}/M_\odot) = 12.79^{+0.34}_{-0.32}$ for the bright sample and $\log(M_\mathrm{vir}/M_\odot) = 12.20^{+0.40}_{-0.30}$ for the faint sample. The errors represent the $16^{\rm th}$ and $84^{\rm th}$ percentiles of the distributions. These frequency distributions are shown as gray histograms in Extended Data Figure \ref{fig:ext4} and are normalized to the total number of bright and faint sources selected, respectively. 

To obtain systems that mimic the MUDF quasars in the sky, we select all pairs separated by a projected physical distance in the range of $400-600$~pkpc, encompassing the projected separation observed for the MUDF pair. The selection leads to a pair density of $\sim 3\times10^{-8}$~cMpc$^{-3}$.
We also include the redshift information by relying on optical rest-frame spectroscopy of the two MUDF quasars. Noting that the actual separation in velocity space for the MUDF pair is $\Delta v = 568 \pm 355~\rm km~s^{-1}$, we require that the selected pairs have $\Delta v \leq 1000~\rm km~s^{-1}$, resulting in a number density of $5.6\times 10^{-9}$~cMpc$^{-3}$. 
Thus, to find at least one pair similar to those in the MUDF, we need to sample a cube volume with a comoving side of $\sim 560$~cMpc, which is $\sim 75$~percent of the comoving side of the cube used in the \texttt{Millenium} simulation. The configuration of the MUDF is thus rare but not highly uncommon in the high-redshift Universe.

\begin{figure}
\centering
\includegraphics[scale=0.5]{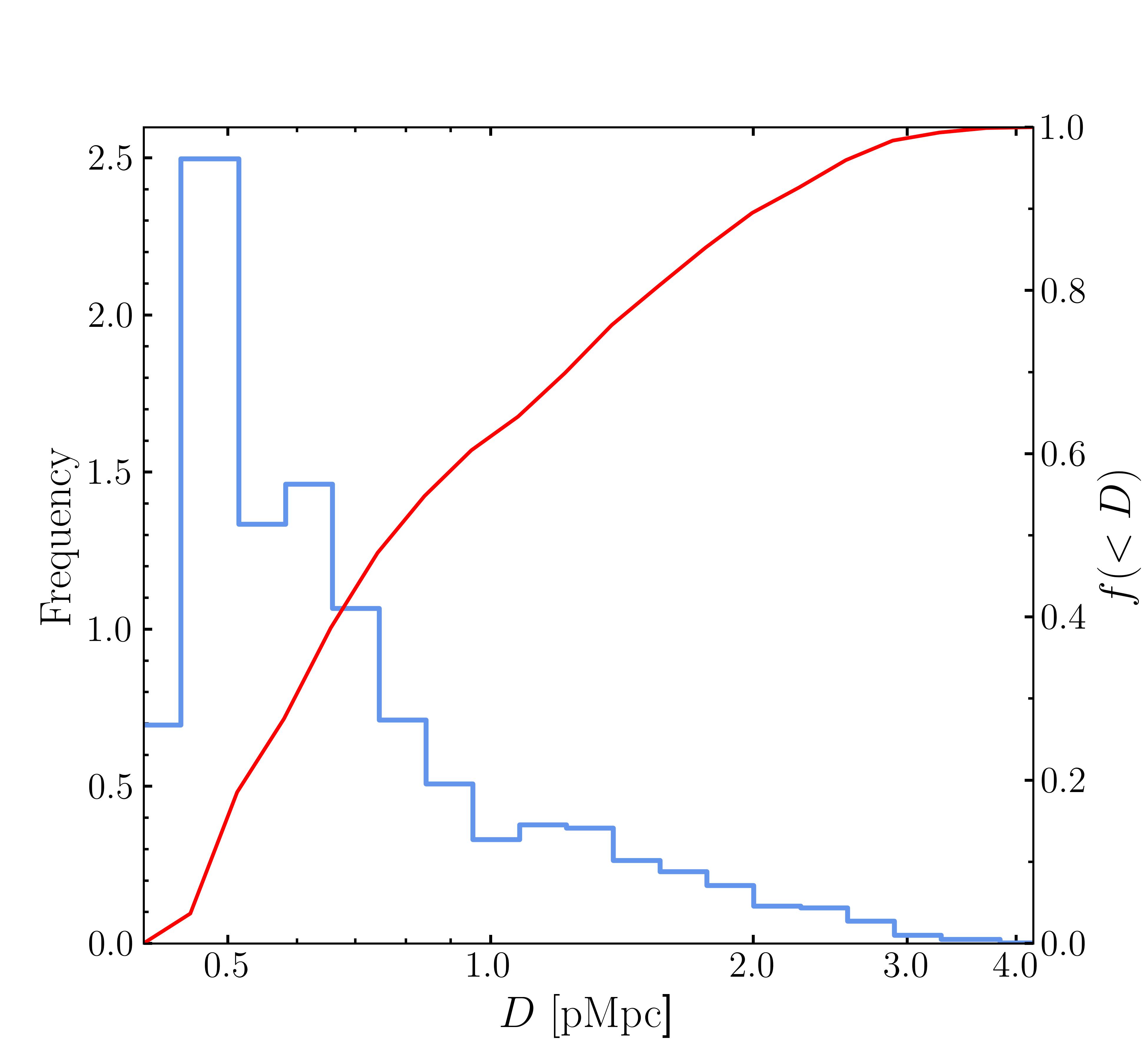}
\caption{\textbf{3D distance distribution of selected pairs.} The frequency distribution of the 3D physical distance between each selected pair (blue histogram) and the corresponding cumulative probability function (red line) are shown. The distribution is normalized to the number of selected pairs. A large fraction of MUDF pair twins are found to be sufficiently close to be interacting in some form.}
\label{fig:ext3}
\end{figure}

The pairs' virial halo mass frequency distributions under this final selection are shown in Extended Data Figure \ref{fig:ext4} (red and blue curves for the bright and faint quasars). The mean values are $\log(M_\mathrm{vir}/M_\odot) = 12.91^{+0.34}_{-0.33}$ for QSO1 and $\log(M_\mathrm{vir}/M_\odot) = 12.25^{+0.46}_{-0.35}$ for QSO2. As above, the errors represent the $16^{\rm th}$ and $84^{\rm th}$ percentiles of the distributions, which we normalized to the total number of bright and faint sources obtained after the final selection. 
The mean masses are slightly larger than those obtained using only the bolometric luminosity selection, as imposing stringent constraints on the physical distance between halos allows us to preferentially select more biased regions than random pairs.
We then derive the underlying 3D physical distance distribution, as shown in Extended Data Figure \ref{fig:ext3}, and the cumulative distribution function (red line); $95$~percent of the systems are closer than $2.5$~pMpc and more than $50$~percent are closer than $800$~pkpc. Thus, a large fraction of pairs in the MUDF configuration are part of the same large-scale structure and could be interacting in some form.

\subsection*{Analysis of hydrodynamic simulations}\label{sm:sims}

With the mass distribution of the MUDF pair in hand, we consider hydrodynamic cosmological simulations to explicitly verify the hypothesis that gaseous structures physically connect systems similar to the MUDF pair twins. 
We consider the IllustrisTNG simulations \cite{Nelson2019}, focusing specifically on TNG100-1, the intermediate periodic simulation box of side length $\sim 100$ pMpc. With a gas particle mass of $1.4\times10^6~M_\odot$ and a dark matter particle mass of $7.5\times10^6~M_\odot$, TNG100-1 balances volume and resolution. Each TNG simulation includes a comprehensive model for galaxy formation and solves the coupled evolution of dark matter, cosmic gas, luminous stars, and supermassive black holes from $z=127$ to $z=0$. The simulation generates several snapshots across cosmic time, and for our analysis, we consider the one at redshift $z=3.28$, similar to the redshift of the MUDF system. 

Using the halo mass distributions of the pairs similar to MUDF obtained from SAM (see Extended Data Figure \ref{fig:ext4}), we select a sample of pairs within TNG100-1 with halo masses matching those of each SAM pair within $0.1$ dex. We require a projected physical distance in the range $400-600$~pkpc and a 3D distance below 5 pMpc, according to the 3D physical distance distribution of the SAM in Extended Data Figure \ref{fig:ext3}. We analyze separately two distinct regimes: pairs with a 3D distance below $1$~pMpc (144 close pairs) and those at a larger distance, above $2$~pMpc (52 distant pairs). 

We calculate the hydrogen density profile for each pair along the direction that connects the halos, considering all gas resolution elements within a cylinder positioned between the two halos. The cylinder's axis corresponds to the line connecting the two halos, and the cylinder's radius is set to $100$~pkpc to encompass potential filamentary structures in between. We verified that this geometry fully encompasses the filaments in nearly the entire sample. After normalizing the length of the cylinder by the 3D physical distance of the pair, we divide it into $15$ uniformly distributed slices to ensure a sufficient sampling of the profile. We verify that alternative slice choices do not affect the results. For each slice, we compute the hydrogen density as the total mass contained in the slice divided by the volume of the slice, assuming a primordial hydrogen fraction $X_\mathrm{H}=0.76$. 

\begin{figure}
\centering
\includegraphics[scale=0.5]{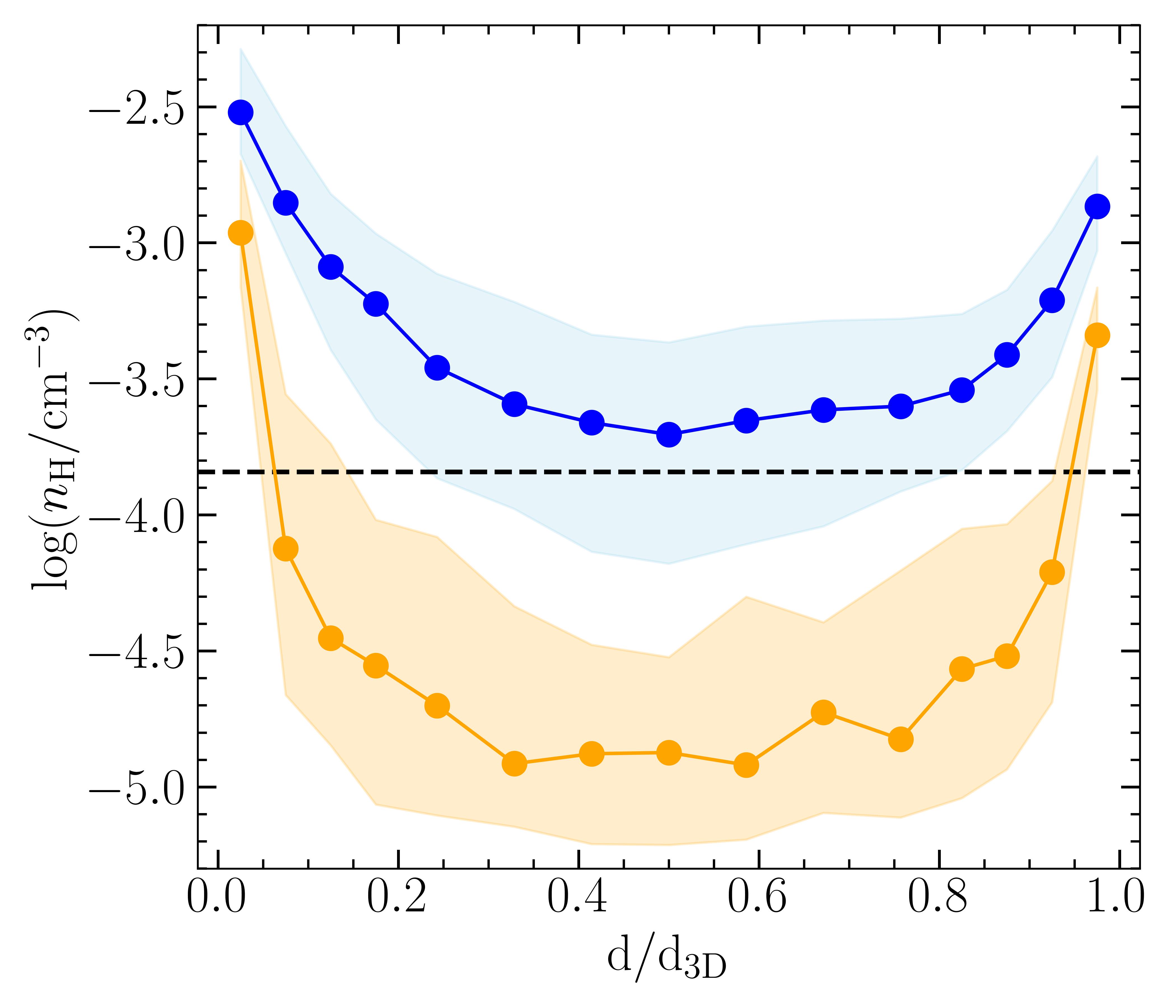}
\caption{\textbf{Median hydrogen density profiles for close and distant pairs.} The median hydrogen density profile along the filament of the selected close and distant pairs separated by a 3D distance of $<1$~pMpc (blue line) and $>2$~pMpc (orange line), are shown.
Both profiles are plotted as a function of the normalized 3D physical distance between the two halos. The shaded regions represent the $16^{\rm th}$ and $84^{\rm th}$ percentiles of the profiles distribution. The black dashed lines represent ten times the hydrogen critical density of the Universe at this redshift, the threshold used to compare the properties of the two subsamples.}
\label{fig:ext5}
\end{figure}

The median density profiles of the pairs, both those with a 3D distance below 1~pMpc and those above 2~pMpc, are shown in Extended Data Figure \ref{fig:ext5}. The sky-blue and orange-shaded regions mark the $16^{\rm th}$ and $84^{\rm th}$ percentiles of the profile distribution. Pairs with a 3D distance below 1~pMpc are typically connected by a denser medium (see, e.g., the Fig. \ref{fig:4}), exhibiting a smooth transition from the CGM to the IGM, with a minimum median hydrogen density value of $\sim 10^{-3.7}$~cm$^{-3}$. In contrast, distant pairs with a 3D distance above 2~pMpc are not typically connected by an identifiable overdense structure and display a steeper radial hydrogen density profile, reaching a minimum median hydrogen density of $\sim 10^{-4.8}$~cm$^{-3}$. We interpret this result as statistical evidence of more overdense filaments connecting the close pairs. To further quantify the occurrence of connecting filaments between the two subsamples, we also calculate the hydrogen density within a cylinder of radius $100$~pkpc in the region ranging from $0.25d_\mathrm{3D}$ to $0.75d_\mathrm{3D}$, focusing solely on the contribution of the filamentary structure and excluding the region associated with the CGM. We determine the fraction of pairs with a filament density value above a threshold of ten times the critical hydrogen density at redshift $\sim 3.22$, which is $\sim 1.4\times 10^{-5}$~cm$^{-3}$. This reference value aligns with the strongest absorbers observed in the \lya\ forest from quasar spectra (e.g., \cite{Rauch1998}). Most systems with a 3D distance below 1~pMpc ($75$~percent) exhibit densities exceeding this threshold, while only $8$~percent of systems with a 3D distance above 2 pMpc exceed this threshold. We thus conclude that the former subsample contains systems truly connected by a dense gaseous filament.

\begin{figure}
\centering
\includegraphics[scale=0.45]{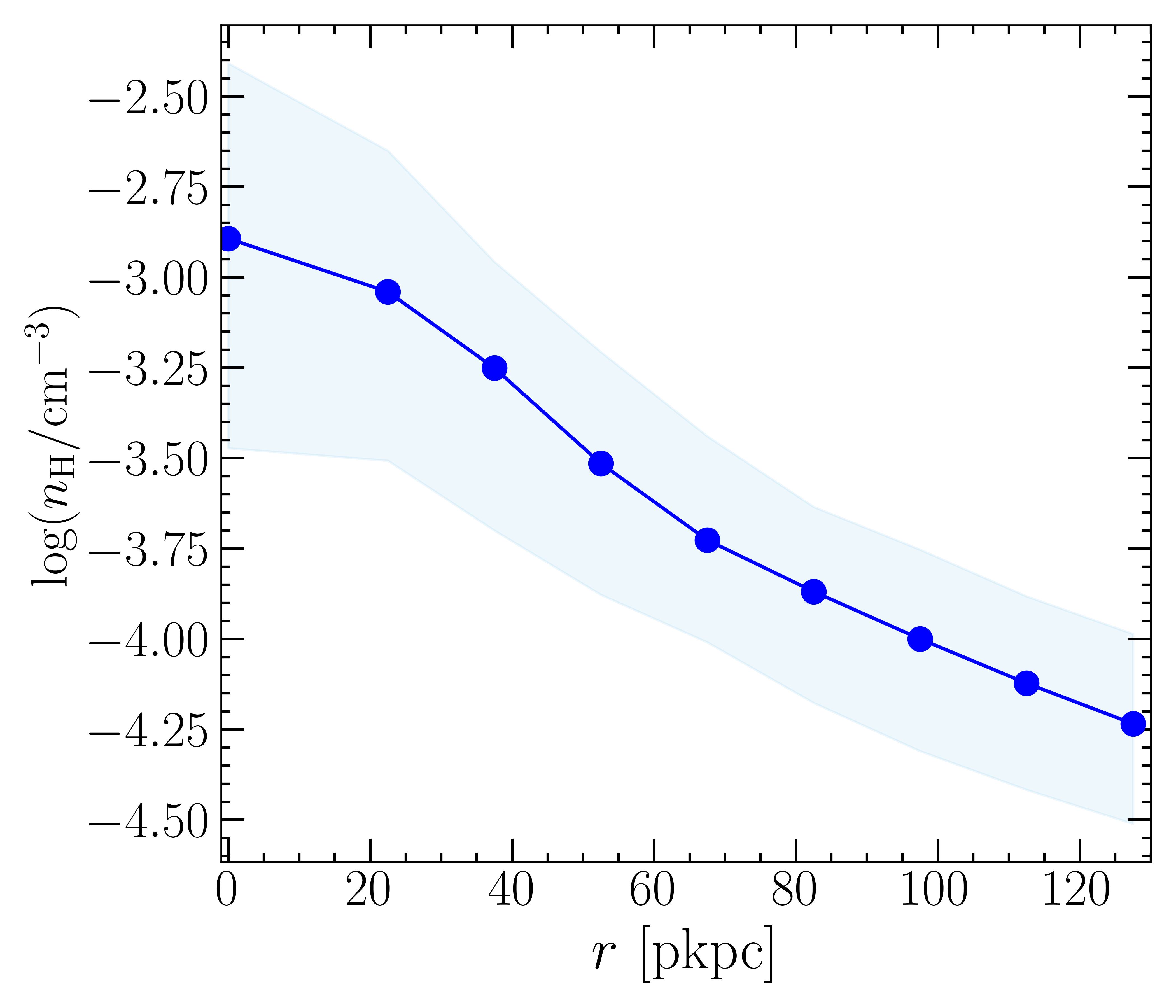}
\caption{\textbf{Transverse median hydrogen density profile for physically connected pairs}. The transverse median hydrogen density profile of the physically connected pairs separated by a 3D distance $<1$~pMpc is shown. The sky-blue colored region represents the $16^{\rm th}$ and $84^{\rm th}$ percentiles of the profiles distribution.}
\label{fig:ext6}
\end{figure}

We also compute the filament's transverse hydrogen profile for the physically connected pairs with a 3D distance below 1~pMpc. After selecting the filamentary structure as described above, we define the direction of the filament's spine, computing the two highest density points below and above $0.5d_\mathrm{3D}$. With this approach, we can account for filaments not perfectly aligned with the axis joining the two halos, as done in our analysis of the transverse surface brightness profile in the MUSE data. Knowing the filament's orientation, we consider a cylinder along the spine direction with a radius of $130$~pkpc, which we divide into different shells with a width of $\Delta r=15$~pkpc. As above, we estimate the hydrogen density as the total gas mass contained in a given shell divided by the volume of the shell, after accounting for the hydrogen fraction $X_\mathrm{H}$. The profile obtained is shown in Extended Data Figure \ref{fig:ext6}. From this result, we can infer that the physically connected pairs have a filamentary structure with a median hydrogen density of $\approx 10^{-2.8}$~cm$^{-3}$ in the densest part along the filament's spine and that the density falls from the center of the filament with an exponential decline radius of $40\pm15$~pkpc. 

To more closely compare the MUSE observations with the results of simulations, we derive the surface brightness maps assuming that the diffuse gas emission originates from recombinations and collisional excitation. For each gas resolution element, we calculate the emissivities from the equations
\begin{equation}
\epsilon_\mathrm{Ly\alpha}^\mathrm{rec} = \frac{h\nu_\mathrm{Ly\alpha}}{4\pi}\alpha_\mathrm{eff}(T)(1-\eta)^2n_\mathrm{H}^2,
\label{eq:rec_Rahm}
\end{equation}
and
\begin{equation}
\epsilon_\mathrm{Ly\alpha}^\mathrm{coll} = \frac{h\nu_\mathrm{Ly\alpha}}{4\pi}\gamma_\mathrm{1s2p}(T)\eta(1-\eta)n_\mathrm{H}^2.
\end{equation}
The emissivities depend on the squared number density of neutral hydrogen, $n_\mathrm{H}$. Recombinations are calculated assuming a case A scenario with the temperature-dependent recombination coefficient $\alpha_\mathrm{eff}(T)$ from Hui \& Gnedin \cite{Hui&Gnedin1997} and the collisional excitation coefficient $\gamma_\mathrm{1s2p}(T)$ from Scholz \& Walters \cite{Scholz&Walters1991}. Assuming ionization equilibrium, the neutral hydrogen fraction, $\eta = n_\mathrm{HI}/n_\mathrm{H}$, is calculated following Appendix A2 in Rahmati et al. \cite{Rahmati2013}, as done within the simulation.
The temperature of each cell gas is calculated assuming a perfect monoatomic gas from the internal energy $u$ given by the simulation, using the relation $T_\mathrm{cell} = (\gamma -1)\mu m_\mathrm{p}u/k_\mathrm{B}$, where $\gamma = 5/3$ and $\mu = 4/(1+3X_\mathrm{H}+4X_\mathrm{H}x_\mathrm{e})$ is the mean molecular weight calculated with the electron abundance $x_\mathrm{e}$ given by the simulation.
We include only gas with densities $n_{H}<0.1~\rm cm^{-3}$, i.e., outside the imposed equation of state.
Due to the presence of the quasars in our observations, we also consider a maximal fluorescence model to test their possible effect on the gas in the filaments, assuming that the ionizing sources are bright enough to fully ionize the surrounding medium. Therefore, we calculate the emissivity of the gas due to \lya\ recombination radiation following a simplified relation where $\eta$ is assumed to be zero in equation (\ref{eq:rec_Rahm}) (see, e.g., de Beer et al. \cite{deBeer2023}).
Finally, using the public package \texttt{Py-SPHViewer} \cite{alejandro_benitez_llambay_2015_21703}, we integrate along the line of sight (assumed as $z$) the total emissivity to obtain the surface brightness images of the selected pairs (see, e.g., the Fig. \ref{fig:4}).

\begin{figure}
\centering
\includegraphics[scale=0.45]{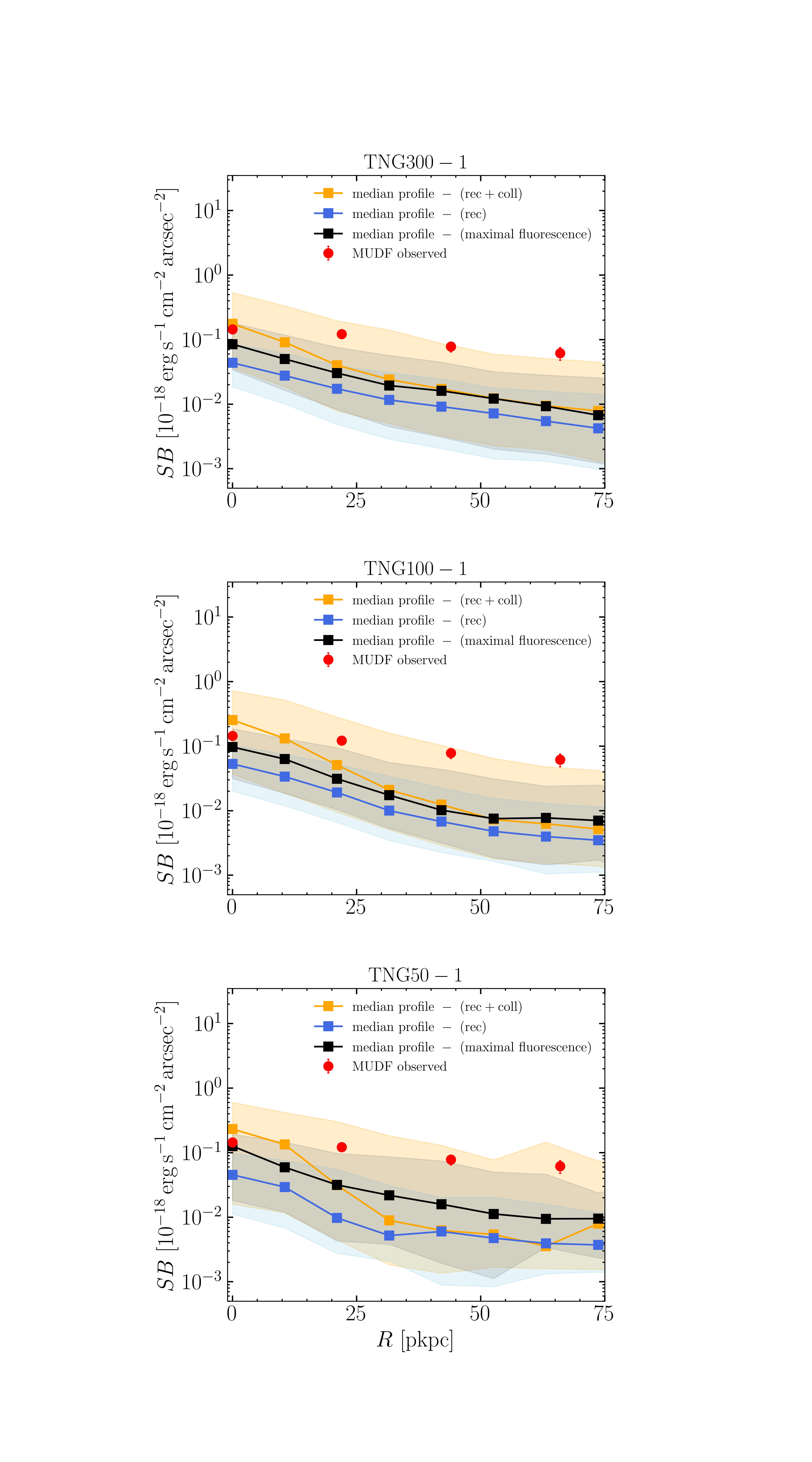}
\caption{\textbf{Transverse surface brightness profiles in the different box sizes of simulation.} \textbf{a-c} The median transverse surface brightness profiles of the selected pairs with a 3D distance below 1 pMpc in TNG300-1, TNG100-1 and TNG50-1, respectively. The blue line represents the emission contribution from recombinations only, while the orange line represents both recombinations and collisional excitations. The emission contribution from the maximal fluorescence model is also shown as a black line. The red points represent the measured data and their errors for the transverse surface brightness profile in MUDF. The shaded regions denote the $16^{\rm th}$ and $84^{\rm th}$ percentiles of the profile distributions.}
\label{fig:ext7}
\end{figure}

Using the same approach as followed in our observations (see also the Extended Data Figure \ref{fig:ext2b}), we measure the transverse surface brightness profile using nine rectangular boxes for each pair up to a distance of $100$~pkpc on either side relative to the direction connecting the two halos. Once again, the boxes can be positioned off-axis to that direction, ensuring the emission peak is at $R=0$. To exclude the contribution from the CGM of the two main halos (but leaving possible contribution from other embedded halos), each box's length encompasses only the projected filament region, defined by maintaining a distance of $0.25d_\mathrm{2D}$ from each halo, which is similar to the transition radius measured in the MUDF.
As both emission processes considered here depend on the density square, we explicitly test the robustness of these predictions as a function of resolution, comparing the results in three boxes of the IllustrisTNG simulation (TNG300-1, TNG100-1, and TNG50-1), covering a range of $\approx 200$ in volume and $\approx 130$ in mass. Applying the pair selection described above for TNG100-1 yields 710 pairs below 1~pMpc in TNG300-1 and 24 in TNG50-1.

Extended Data Figure \ref{fig:ext7} compares the median emission profile from both recombinations and collisional excitations (orange line), as well as considering only recombinations (blue line), along with the $16^{\rm th}$ and $84^{\rm th}$ percentiles. We observe that, on average, the result is not strongly sensitive at the different resolutions of TNG, implying that the typical densities within the mildly overdense filaments are reasonably converged at these scales, a result also found in simulations of the Ly$\alpha$ forest.
We also observe that the maximal fluorescence model produces a median surface brightness profile that agrees remarkably well, within a mean factor of $\approx 2$, with the recombination model. 
Thus, the simulations predict that the mostly optically thin filaments have temperature-dependent coefficients and ionization fractions close to the maximal fluorescence conditions, a result similar to the simulation predictions in \cite{Kollmeier2010}, where the surface brightness of the faintest pixels originates from low column-density material that is already highly ionized and emitting near its maximum.
This analysis concludes that the surface brightness maps are generally robust relative to the assumptions made.

When comparing the predicted profiles with the observational data points measured in the MUDF up to $\approx 70$ pkpc, where the measurements exceed the detection limit of the NB image, we observe that the maximal fluorescence and the recombination models lie below the observed surface brightness.
This indicates that the densities predicted by the simulations cannot be too high compared to real values as, otherwise, the simulated profiles would exceed the observed ones. Moreover, observations and simulations can be brought into agreement by increasing recombination radiation by a factor of $\approx 9$ in the surface brightness, i.e., requiring an increase in density by a factor of not more than $\approx 3$.
Hence, the simulated densities cannot be much lower than the true values.
Such a boost should also be considered a maximum correction that must be applied due to the presence of additional photons from collisions. Indeed, when including collisional excitations, despite the more uncertain nature of this calculation due to the high sensitivity to temperature, the profile shifts upward, especially in the inner $\approx 25~$pkpc.
We also note that the $0.1~\rm cm^{-3}$ cut is quite stringent, as we explicitly tested that the denser and more neutral gas around $\approx 0.1-0.3~\rm cm^{-3}$ is a significant source of photons. Including that phase, we observe a good agreement between the surface brightness level predicted by the simulations and the MUDF filament, with both profiles lying at $\approx 10^{-19}~\rm erg~s^{-1}~cm^{-2}~arcsec^{-2}$ in the range $R\approx 20-40$~pkpc, reducing the flattening at larger radius. Our analysis also neglects scattering processes that can redistribute photons from regions of high to low surface brightness. Byrohl \& Nelson \cite{Byrohl&Nelson2023} quantify the effects of radiative transfer in the TNG simulation, and at the surface brightness level observed in the MUDF filament, the boost factor is approximately 2-3. 
Hence, discrepancies are not particularly concerning. Moreover, we observe that this analysis does not require the introduction of large clumping factors to explain the SB levels, suggesting that the already mostly ionized, optically thin filaments have a simpler density distribution, reasonably captured by the simulations as tested above. High-density clumps, as required in the bright large \lya\ nebulae, would produce surface brightness values that exceed the observed ones also by two orders of magnitudes (\cite{Cantalupo2014}).
Overall, this analysis implies a satisfactory agreement between the density predicted in the cold dark matter model and what is observed.

Furthermore, there is no special reason why the MUDF filament should align with the distribution median. Considering this aspect, we searched among the simulated surface brightness profiles for a pair that resembles the MUDF system more closely. One such MUDF twin is shown in Fig. \ref{fig:4}, where we see a transverse profile that matches the observations. 
Hence, filaments with observed characteristics comparable to the MUDF exist in the cold dark matter paradigm, and our study paves the way for further quantitative analysis of the properties of the cosmic web within our cosmological model.

\backmatter

\subsection*{Data Availability}
The VLT data used in this work are available from the European Southern Observatory archive 
\href{https://archive.eso.org/}{https://archive.eso.org/} either as raw data or phase 3 data products \cite{Fumagalli2023}.

\subsection*{Acknowledgments}
This project has received funding from the European Research Council (ERC) under the European Union's Horizon 2020 research and innovation program (grant agreement No 757535 and No 101026328), by Fondazione Cariplo (grant No 2018-2329) and is supported by the Italian Ministry for Universities and Research (MUR) program ``Dipartimenti di Eccellenza 2023-2027'', within the framework of the activities of the Centro Bicocca di Cosmologia Quantitativa (BiCoQ). DIV acknowledges financial support provided under the European Union’s H2020 ERC Consolidator Grant “Binary Massive Black Hole Astrophysics” (B Massive, Grant Agreement: 818691). SC and AT gratefully acknowledge support from the European Research Council (ERC) under the European Union’s Horizon 2020 Research and Innovation programme grant agreement No 864361. PD acknowledges support from the NWO grant 016.VIDI.189.162 (``ODIN") and warmly thanks the European Commission's and University of Groningen's CO-FUND Rosalind Franklin program. SB acknowledges support from the Spanish Ministerio de Ciencia e Innovación through project PID2021-124243NB-C21. This research made use of Astropy, a community-developed core Python package for Astronomy (\cite{Astropy2013, Astropy2018, Astropy2022}, NumPy (\cite{Harris2020}), SciPy (\cite{Virtanen2020}), Matplotlib (\cite{Hunter2007}).

\subsection*{Author contributions}
DT analyzed the observations and was the main author of the manuscript. MiFu coordinated the MUDF program, participated in the data analysis, and co-authored the manuscript. MaFo reduced and analyzed the observations and participated in the analysis and manuscript writing. ABL contributed to the simulation analysis and DIV created and provided the SAM lightcone, and contributed to the analysis. All co-authors participated in preparing the manuscript.

\subsection*{Competing interests}
The authors declare no competing interests.


\bibliography{sn-bibliography}

\end{document}